\documentclass[fleqn,usenatbib]{mnras}
\pdfoutput=1

\usepackage{amsmath}	
\usepackage{txfonts}
\usepackage{CJKutf8}

\usepackage[T1]{fontenc}
\usepackage{ae,aecompl}

\usepackage{graphicx}	
\usepackage{amsmath}	
\usepackage{amssymb}	
\usepackage{multirow}

\usepackage{color}

\newcommand{\fesc}{\ifmmode{f_{\rm esc}}\else{$f_{\rm esc}$}\fi}
\newcommand{\fescs}{\ifmmode{f_{\rm esc}^\star}\else{$f_{\rm esc}^\star$}\fi}
\newcommand{\kms}{\ifmmode{{\;\rm km~s^{-1}}}\else{km~s$^{-1}$}\fi}
\newcommand{\fgas}{\ifmmode{{f_{\rm gas}}}\else{$f_{\rm gas}$}\fi}
\newcommand{\cubecm}{\ifmmode{{\rm cm^{-3}}}\else{cm$^{-3}$}\fi}
\newcommand{\ztwo}{\ifmmode{{\rm [Z_2/H]}}\else{[Z$_2$/H]}\fi}
\newcommand{\zthree}{\ifmmode{{\rm [Z_3/H]}}\else{[Z$_3$/H]}\fi}
\newcommand{\lsim}{\lower0.3em\hbox{$\,\buildrel <\over\sim\,$}}
\newcommand{\gsim}{\lower0.3em\hbox{$\,\buildrel >\over\sim\,$}}

\newcommand{\eavg}{\ifmmode{\langle E_\gamma \rangle}\else{$\langle E_\gamma \rangle$}\fi}

\newcommand{\Ms}{\ifmmode{\textrm{M}_\odot}\else{M$_\odot$}\fi}
\newcommand{\vrms}{\ifmmode{v_{\rm rms}}\else{$v_{\rm rms}$}\fi}

\newcommand{\hh}{H$_2$}

\newcommand{\tvir}{\ifmmode{T_{\rm{vir}}}\else{$T_{\rm{vir}}$}\fi}
\newcommand{\mvir}{\ifmmode{M_{\rm{vir}}}\else{$M_{\rm{vir}}$}\fi}
\newcommand{\rvir}{\ifmmode{r_{\rm{vir}}}\else{$r_{\rm{vir}}$}\fi}

\newcommand{\lya}{Ly$\alpha$}
\newcommand{\jj}{\ifmmode{J_{21}}\else{$J_{21}$}\fi}
\newcommand{\flw}{\ifmmode{F_{LW}}\else{$F_{LW}$}\fi}
\newcommand{\kph}{\ifmmode{k_{\rm ph}}\else{$k_{\rm ph}$}\fi}

\newcommand{\zsun}{\ifmmode{\rm\,Z_\odot}\else{$\rm\,Z_\odot$}\fi}

\newcommand\unit[1]{\; \textrm{#1}}

\title[Ly$\alpha$ trapping in direct collapse black hole formation]
{On the effect of Lyman alpha trapping during the initial collapse of
  massive black hole seeds}

\author[Q. Ge and J.~H. Wise]{
Qi Ge (\begin{CJK}{UTF8}{gbsn}葛琦\end{CJK})$^{1}$\thanks{E-mail: qge30@gatech.edu} and
John H. Wise$^{1}$\thanks{E-mail: jwise@physics.gatech.edu}
\\
$^{1}$Center for Relativistic Astrophysics, School of Physics, Georgia Institute of Technology, Atlanta, GA 30332, USA
}

\pubyear{2017}

\begin{document}
\label{firstpage}
\pagerange{\pageref{firstpage}--\pageref{lastpage}}
\maketitle

\begin{abstract}
  One viable seeding mechanism for supermassive black holes is the
  direct gaseous collapse route in pre-galactic dark matter halos,
  producing objects on the order of $10^4 - 10^6$ solar masses.  These
  events occur when the gas is prevented from cooling below
  $10^4$~K that requires a metal-free and relatively H$_2$-free
  medium.  The initial collapse cools through atomic hydrogen
  transitions, but the gas becomes optically thick to the cooling
  radiation at high densities.  We explore the effects of
  Lyman-$\alpha$ trapping in such a collapsing system with a suite of
  Monte Carlo radiation transport calculations in uniform density and
  isotropic cases that are based from a cosmological simulation.  Our
  method includes both non-coherent scattering and two-photon line
  cooling.  We find that Lyman-$\alpha$ radiation is marginally
  trapped in the parsec-scale gravitationally unstable central cloud,
  allowing the temperature to increase to 50,000~K at a number density
  of $3 \times 10^4$~cm$^{-3}$ and increasing the Jeans mass by a
  factor of five.  The effective equation of state changes from
  isothermal at low densities to have an adiabatic index of 4/3 around
  the temperature maximum and then slowly retreats back to isothermal
  at higher densities.  Our results suggest that Lyman-$\alpha$
  trapping delays the initial collapse by raising the Jeans mass.
  Afterward the high density core cools back to $10^4$~K that is
  surrounded by a warm envelope whose inward pressure may alter the
  fragmentation scales at high densities.
\end{abstract}


\begin{keywords}
  cosmology: dark ages -- supermassive black holes -- radiative
  transfer
\end{keywords}



\section{Introduction}
\label{sec:intro}
Observations of bright quasars at redshifts $z \gtrsim 6$ indicate
that supermassive black holes (SMBHs) with masses over $10^9~\Ms$ form
within the first billion years after the Big Bang. \citep{Fan, Willot,
  Mortlock, Wu}. These SMBHs are expected to form by seeding
mechanisms that can be categorized into three classifications: the
growth of massive metal-free (Population III; Pop III) stellar remnants
\citep{Madau, Volonteri}, collapse of dense stellar clusters
\citep{Davies} and a direct collapse of a gaseous metal-free cloud
\citep{Bromm, Wise, Begelman, Volonteri2}. Light BH seeds from Pop III
stars will have a difficult time growing at the Eddington limit into
the observed high-redshift quasars because of the warm and diffuse
medium left behind by its progenitor star and the limited period
between their formation and redshift~6 \citep{Johnson, Alvarez, Jeon};
however, hyper-Eddington accretion may overcome this barrier
\citep{Alexander, Inayoshi}. Furthermore after a BH merger, the kick
velocity of the resulting BH is most likely greater than the escape
velocity of their host dark matter halos \citep{Herrmann, Micic}.

In the direct collapse scenario, which is the focus of this work,
halos with a virial temperature $T_{\rm vir} \simeq 10^4 \unit{K}$
($M_{\rm vir} \gtrsim 10^8~\Ms$ at $z \sim 10$), known as atomic
cooling halos, that are chemically pristine and have a very low
molecular hydrogen density can catastrophically collapse \citep{Rees,
  White}. This happens at such a virial temperature because the atomic
hydrogen ionization and collisional de-excitation rates increase by
several orders of magnitude. The general criterion for a rapid gaseous
collapse is that the gas cooling time is less than the free-fall
time. It is thought that the massive baryon cloud collapses
monolithically and isothermally without fragmentation in a
Lyman-Werner (LW) background\footnote{$J_{21}$ is the background
  specific intensity in units of $10^{-21} \unit{erg} \unit{s}^{-1}$
  $\unit{cm}^{-2} \unit{Hz}^{-1} \unit{sr}^{-1}$ at the Lyman limit
  (13.6~eV).}  $J_{21} > J_{\rm crit} \simeq 10^3$ given a $10^5
\unit{K}$ blackbody spectral shape \citep[e.g.][]{Omukai, Shang, WG12,
  Agarwal15, Glover15}.

Numerical simulations have indeed shown that fragmentation is
suppressed when H$_2$ cooling is absent \citep{Bromm, Regan}. The
Jeans mass
\begin{equation}
  M_{\rm J} \simeq 10^{5.5}~\Ms 
  \left(\frac{T}{10^4 \unit{K}} \right)^{3/2}
  \left(\frac{n}{10^{4} \unit{cm}^{-3}}\right)^{-1/2},
\end{equation}
determines the approximate fragmentation mass scale during the
collapse and is a key characteristic quantity to follow during this
phase. The gas temperature $T$ highly depends on whether radiative
cooling is efficient, in particular H$_2$ in a metal-free gas. When it
is efficient, the gas can cool down to $T \sim 300 \unit{K}$,
corresponding to $M_{\rm J} \sim 10^3~\Ms$, implying that the cloud
will form massive Pop III stars.  Prior to reionization, the LW
background is not sufficiently high to affect all atomic cooling halos
\citep{Visbal14_LWB}, but there is a small possibility that such a
pre-galactic halo has a nearby neighboring galaxy that boosts the
impinging LW radiation above $J_{\rm crit}$ \citep[e.g.][]{Dijkstra08,
  Agarwal14, Visbal14_Pairs, Regan2, Regan17}.  Without \hh{} cooling,
atomic hydrogen transitions allow the gas to cool to 8000~K but no
further, resulting in a central Jeans mass $M_{\rm J} \sim 10^5 -
10^6~\Ms$, that has the possibility of collapsing into a dense stellar
cluster or a supermassive star, ultimately producing a massive black
hole on the order of $M \simeq 10^4 - 10^6~\Ms$.

Coherent scattering properties of Ly$\alpha$ photons have been studied
for decades, initially focusing on analytical treatments of radiation
scattering \citep{Unno, Hummer, Adams} and the Eddington approximation
\citep{Harrington, Neufeld, Loeb2}. More recent studies utilize Monte
Carlo methods in several different scenarios: the emerging spectrum
from an isothermal homogeneous medium with plane-parallel or spherical
symmetry \citep{Ahn, Zheng}, an isotropic velocity field
\citep{Dijkstra}, a density gradient field \citep{Barnes}, dust
absorption and re-emission \citep{Verhamme}, Ly$\alpha$ radiative
transfer shells model \citep{Gronke} and transmission through the
intergalactic medium \citep{Laursen}.

Ly$\alpha$ trapping has been considered to be an important impact
factor on the formation of direct collapse black holes \citep{Spaans,
  Latif, Yajima2}. The scattering of photons in the dense
optically-thick core will limit gas cooling and possibly increase the
temperature, leading to a higher Jeans mass. Furthermore, radiation
trapping leads to a breakdown of the Eddington limit, making
hyper-Eddington accretion onto BHs a possibility
\citep{Inayoshi}. This transition from an optically-thin cooling limit
to an optically-thick medium has been previously approximated with a
polytropic equation of state, derived in spherical symmetry, that
evolves from isothermal to adiabatic in a range $n = 1-10^5
\unit{cm}^{-3}$ \citep{Spaans}. In this model, the adiabatic behavior
at high densities will keep the gas nearly H$_2$ free during the
collapse.

The primary aim of this work is to examine the thermodynamics of the
direct collapse to a massive BH seed in an atomic cooling halo.  We
first construct a radiative cooling model that includes the effects of
\lya{} trapping that allows us to explore under what conditions the
gas deviates from isothermal.  Then we perform a cosmological
simulation focusing on an atomic cooling halo from which we extract
radial profiles and then perform a suite of Monte Carlo \lya{}
radiation transport calculations in various idealized cases.  From
these results, we estimate the effective equation of state of the
collapsing system, shedding light on the expected mass scale of the
central object.  The fragmentation scale and final outcome of such a
primordial collapse are still open questions, and we aim to edge
closer to their answers by including another key physical process in
its initial collapse.

The rest of the paper is organized as follows. In \S\ref{sec:2} we
describe our radiative cooling model, including an approximate model
of Ly$\alpha$ trapping whose details are left for the Appendix, the
cosmological simulation, and the \lya{} radiation transport
calculation. In \S\ref{sec:3}, we present the results of our radiative
cooling rates with \lya{} trapping and a suite of Monte Carlo
calculations, focusing on the effects of Ly$\alpha$ trapping on the
thermodynamics of the central collapse. In \S\ref{sec:4}, we conclude
and discuss the impact of Ly$\alpha$ trapping especially regarding to
fragmentation and discuss the limitations of our method with future
directions on resolving the full evolutionary sequence to a massive BH
seed.


\section{Methods}
\label{sec:2}
We investigate the thermal evolution of a collapsing metal-free gas
cloud with two methods.  First, we calculate a cooling rates as a
function of temperature, i.e. the cooling curve, when the effects of
\lya{} trapping are included.  Second, we build upon these results by
running a cosmological simulation that focuses on a halo that
potentially hosts a direct collapse black hole.  We then post-process
several of these snapshots in a \lya{} radiation transport
calculation, where we quantify the propagation of such photons and how
the cooling rates deviate from the optically-thin approximation.

\subsection{Radiative cooling with \lya{} radiation trapping}
\label{sec:cooling}

We first modify the primordial gas cooling curve to include the
effects of \lya{} radiation trapping.  This model is similar to
previous one-zone models and chemical networks \citep[e.g.][]{Cen,
  Omukai, Schleicher, Shang, Glover15}.  Generally, the thermal
evolution of one-zone models in a free-fall collapse lends for a
convenient check for possible fragmentation mass scales.  These models
consider a full chemical network to calculate the cooling rates, and
we initially approach the problem of including \lya{} trapping by
inspecting how it modifies the cooling curve.  Here we adopt the
radiative cooling rates from \citet{Cen} as a basis that include
collision ionization and excitation, recombination, bremsstrahlung,
and Compton cooling from a primordial gas.  We also include molecular
hydrogen cooling, using the rates from \citet{Glover08}, however we
consider a strong LW radiation background that suppresses its efficacy
when $J_{21} \gtrsim 10^3$.  We calculate the cooling curve in a
temperature range of $\log (T/\mathrm{K}) = 2-6$ and number density
range of $\log (n/\mathrm{cm}^{-3}) = 2-9$.

We now review the resonance scattering properties of \lya{} radiative
transfer in a pure hydrogen gas to demonstrate how \lya{} trapping
occurs in an optically-thick medium.  While scattering between
hydrogen atoms, a single \lya{} photon will undergo a frequency change
from Doppler effects.  As is convention and for convenience, we refer
to the frequency in terms of the Doppler width of the line, $x \equiv
(\nu - \nu_0) / \Delta\nu_{\rm D}$, arising from the thermal
velocities of the atoms.  Here $\Delta\nu_{\rm D} = \nu_0 (2k_{\rm
  B}T/m_{\rm p}c^2)^{1/2}$ is the Doppler width; $\nu_0 = 2.466 \times
10^{15} \unit{Hz}$ is the rest-frame frequency of the \lya{}
transition, and $k_{\rm B}$ and $m_{\rm p}$ are the Boltzmann constant
and the proton mass, respectively.  For a zero temperature gas, the
optical depth in the \lya{} line follows a Lorentzian profile with
respect to frequency $x$.  However, when a Maxwellian velocity
distribution is considered, the \lya{} line optical depth transforms
into a Voigt profile
\begin{align}
  \tau_{\nu}^{\rm Voigt} = &
  \frac{\sqrt{\pi} e^2}{m_{\rm e} c}
  N_{\rm HI} f_{12} \; \times\nonumber\\
  &\int \frac{dv}{b} e^{-v^2/b^2}
  \frac{4\gamma_{12}} {16{\pi}^2 [\nu-(1-v/c)\nu_0]^2+\gamma_{12}^2},
  \label{eq:1}
\end{align}
where $f_{12} = 0.4162$ is the \lya{} oscillator strength, $N_{\rm
  HI}$ is the neutral hydrogen column density, and the hydrogen
velocity dispersion is parameterized as the Doppler parameter
\begin{equation}
  b \equiv \sqrt{\frac{2}{3} \langle v^2 \rangle} = 
  \sqrt{ \frac{2k_{\rm B}T}{m_{\rm p}} }.
\end{equation}
Finally, the natural width $\gamma_{12} = A_{12}$ of the line is
related to the Einstein A-coefficient $A_{12} = 6.24 \times 10^8
\unit{s}^{-1}$.  The Voigt profile can be difficult to integrate
analytically, however it can be approximated by introducing the Voigt
parameter $a \equiv A_{12}/4\pi\Delta\nu_{\rm D}$, allowing us to
rewrite the integral as
\begin{equation}
  H(a,x) \equiv \frac{\tau_{x}}{\tau_0} =
  \frac{a}{\pi} \int^{\infty}_{-\infty} dy
  \frac{e^{-y^2}}{(y-x)^2+a^2} \simeq
  \begin{cases} 
    \; e^{-x^2} & \mathrm{(core)}\\ 
    \; a / (x^2 \sqrt{\pi}) & \mathrm{(wing)}\\
  \end{cases},
  \label{eqn:voigt}
\end{equation}
defining $y \equiv v/b$.  Here $\tau_0$ represents the optical depth
at the line center.  Different from radiation transport of continuum
photons, \lya{} photons experience a very short mean free path and is
shortly re-emitted after absorption.  Such a physical system requires
the inclusion of the scattering term in the radiative transfer
equation when following the evolution and morphology of the \lya{}
radiation field.  For instance, the rate of escaping photons from some
system depends on the mean number of scatterings and the associated
frequency shifts, where an escaping photon will likely be in the wing
of the \lya{} line where optical depths are minimal.

We utilize a simplified scattering and trapping model for \lya{}
radiation transport in our cooling rate calculation.  In this model,
the photons that are generated from recombination and collisional
de-excitation are not assumed to escape the system.  They can be
trapped if the optical depth is sufficiently high, suppressing any
cooling.  We approximate the effective cooling rate by calculating the
average number of scatterings that photons experience before they
shift into the wing part of the line profile when they escape and cool
the system.  For the interested reader, more details about \lya{}
radiation transfer can be found in Appendix \ref{appendix}.

\lya{} radiation is mainly generated by two mechanisms: recombination
and collisional excitation.  Hydrogen will be ionized at $T \gtrsim
10^4 \unit{K}$ after either being photo- or shock-heated.  For
recombination, a fraction of the captured free electrons will decay
into the ground state through a cascade, producing a \lya{} photon in
the process.  The emissivity of recombination is
\begin{equation}
  \label{eqn:rec}
  \eta^{\rm rec} = f_\alpha \alpha_B h\nu_\alpha n_{\rm e} n_{\rm HII},
\end{equation}
where $f_\alpha$ denotes the ratio of \lya{} photons generated from
case B recombinations, and $\alpha_{\rm B}$ is the case B
recombination rate coefficient.  We take $f_\alpha \simeq 0.68$ as a
constant because it is only weakly dependent on temperature
\citep{Osterbrock}.  The second process includes a collisional
excitation that occurs when an electron decays into the ground state,
producing a \lya{} photon.  The de-excitation coefficient is $A_\alpha
= 3.7 \times 10^{-17} \exp(-h\nu_\alpha/kT) T^{-1/2}$
\citep{Osterbrock} and the associated emissivity is
\begin{equation}
  \label{eqn:dcol}
  \eta^{\rm dcol} = A_\alpha n_{\rm e} n_{\rm HI}.
\end{equation}
In a pure hydrogen gas with a low ionization state, the intrinsic
\lya{} emissivity can be approximated as $\eta^{\rm src} \equiv
\eta^{\rm rec} + \eta^{\rm dcol}$.  At temperatures $T = 10^3 - 10^4
\unit{K}$, the high value of the Einstein A-coefficient $A_{12}$
results in the emissivity being dominated by spontaneous radiation.

At higher densities ($n \gtrsim 10^6 \unit{cm}^{-3}$), the two-photon
process (2s $\rightarrow$ 1s) becomes one of the dominant coolants,
even though its Einstein A-coefficient $A_{\rm 2s-1s} = 8.23
\unit{s}^{-1}$ \citep{Omukai} is significantly smaller than \lya{},
because its radiation is optically thin, especially when the \lya{}
(2p $\rightarrow$ 1s) photons are trapped \citep{Schleicher,
  Johnson2}.  Also in dense gas, H$^-$ cooling through the free-bound
transition ($\textrm{H} + \textrm{e}^- \rightarrow \textrm{H}^- +
\gamma$) becomes important and will emit and scatter \lya{} photons.
However, these transitions are insignificant on the level of $10^{-5}$
with respect to the collisional de-excitation channel.  We can compare
the scattering cross-section of photo-detachment in hydrogen to the
two-photon process, both of which can interfere with typical
spontaneous emission scattering events.  The cross-section of the
two-photon emission is $\sim 10^{-10}$ of the H$^-$ photo-detachment
cross-section.  However, the typical H$^-$ abundance is $\ll 10^{-10}$
when $n < 10^{17}~\cubecm$ \citep{Van_Borm}, and from this low
abundance of H$^-$, we can conclude that photo-detachment processes
can be neglected at the densities explored in this study
\citep[however see][]{Johnson17}.  This assumption will break down at
higher densities $n \gsim 10^{15}~\cubecm$ when both H$^-$ and H$_2$
become abundant \citep{Omukai, Van_Borm}.  However, the free-fall time
is extremely short at these times, and it is unclear whether \lya{}
trapping will play a role during this stage, warranting further work
that is outside the scope of this paper.


\subsection{Cosmological simulation setup}

To determine the impact of \lya{} trapping on the initial collapse of
an atomic cooling halo, we perform a cosmological simulation using
radiative cooling rates calculated in optically-thin limit.  We use a
zoom-in simulation with the adaptive mesh refinement (AMR) code {\sc
  Enzo} \citep{Enzo}, which utilizes an $N$-body particle-mesh solver
for the dynamics of dark matter particles and an piecewise parabolic
Eulerian method for the hydrodynamics \citep{Colella, Bryan_PPM}.

This optically-thin simulation is the basis for our Monte Carlo
calculations of \lya{} radiation transport that are fully described in
the following section.  The initial conditions are generated with {\sc
  MUSIC} \citep{MUSIC} in a comoving volume of (1 Mpc)$^3$ at redshift
$z = 500$.  We consider the following cosmological parameters that are
consistent with the WMAP 9-year results: $\Omega_{\rm DM}=0.235$,
$\Omega_{\Lambda}=0.7185$, $\Omega_{\rm b} h^2 = 0.02256$,
$\sigma_8=0.820$, $n_{\rm s}=0.9710$, $h=0.697$, where the variables
have their typical definitions \citep{Hinshaw}.  The differences
between the WMAP9 and latest Planck parameters \citep{Planck2015}
only has minimal timing impacts on structure formation and within
their uncertainties.

We first perform a pathfinder, low-resolution $64^3$ dark matter
simulation to locate the most massive halo in the volume at $z=9$,
using the HOP halo finding algorithm \citep{Eisenstein}.  Then we
resimulate the volume with a zoom-in setup that has the same
large-scale modes but with higher resolution and baryons.  In this
setup, we use a base AMR grid with $256^3$ particles and cells that is
supplemented with two nested grids, centered on the location of the
most massive halo at $z=9$.  These nested grids are static in the AMR
hierarchy.  The innermost grid has a DM mass resolution of 27.3~\Ms{}
($1024^3$ effective resolution) that is 64 times finer than the top
grid.  The simulation uses up to 20 levels of AMR refinement,
corresponding to a maximal comoving resolution of 0.03 pc.  We refine
the grid on baryon and DM overdensities when they exceed $3 \times
2^{-0.3l}$, where $l$ is the AMR level.  The negative exponent results
in the simulation being super-Lagrangian focusing more resolution at
higher densities.  In addition, the local Jeans length is always
resolved by at least four cells to avoid artificial fragmentation
\citep{Truelove97}.

We consider a chemical network of nine primordial species (H, H$^+$,
He, He$^+$, He$^{++}$, H$^-$, H$_2^+$, H$_2$ and e$^-$) to evolve
their abundance in non-equilibrium \citep{Anninos97, Abel_Species}
with the \hh{} rates from \citet{Glover08}.  We neglect any metal
enrichment because the direct collapse formation scenario requires
that the gas to be warm ($\gtrsim 5000 \unit{K}$) to avoid cooling and
fragmentation.  Thus to focus on this scenario, we consider only
primordial cooling and apply a Lyman-Werner radiation background with
an intensity of $J_{21} = 10^5$ without any self-shielding effects.
We note that this value of $J_{21}$ is artificially high that requires
a very close ($\lesssim 1 \unit{kpc}$) and luminous radiation source,
but we apply such an intense background to remove any effects from
\hh{} cooling in order to focus on \lya{} radiation trapping.

\subsection{Monte Carlo Radiative Transfer} 
\label{sec:rt}


Our main results on the effects of \lya{} trapping originate from
post-processing the most massive halo in the optically-thin
cosmological simulation with a suite of Monte Carlo radiation
transport calculations.  We consider three cases:
\begin{enumerate}
\item A {\bf uniform density case} with hydrogen number densities
  ranging from $10^5 \unit{cm}^{-3}$ to $10^9 \unit{cm}^{-3}$ where
  the photons are propagated for $10^8 \unit{s}$, corresponding to a
  light-crossing time of 1~pc, approximately the radius of the
  Jeans unstable central gas cloud in an atomic cooling halo.
\item A {\bf time-independent isotropic case} whose radial properties
  are derived from the collapse halo in the cosmological simulation at
  its final time, when the maximum density is $3 \times 10^{11}
  \unit{cm}^{-3}$.  This calculation is also integrated for $10^8
  \unit{s}$.
\item A {\bf time-dependent isotropic case} extends the static case,
  where we allow the cloud to contract.  We take the radial averages
  from six outputs, whose maximum number densities range from $3
  \times 10^7 \unit{cm}^{-3}$ to $3 \times 10^{11} \unit{cm}^{-3}$
  with each output having maximum densities approximately an order of
  magnitude apart.  The output times are 65, 255, 1,100, 4,000, and
  12,600~years before the final output.
\end{enumerate}
This treatment builds upon our optically-thick adjustments to the
cooling curve, or equivalently altering the equation of state, and its
application to a cosmological simulation.  We extract the pertinent
time-dependent radially averaged gas properties, such as density,
temperature, and ionization fraction, from the most massive halo as it
is catastrophically collapsing.  We do not extract the velocity
information, but we consider three different cases: a static medium,
radial infall, and solid body rotation.  For the latter two cases, we
explore two different velocities, 1 and 5~\kms, corresponding to 10\%
and 50\% of the sound speed $c_{\rm s}$ for a $T = 10^4\unit{K}$ gas,
and is consistent with velocities found in cosmological simulations of
atomic cooling halos \citep[e.g.][]{Wise07}.

We post-process these data to estimate the evolution of the \lya{}
radiation field during the collapse.  We base our radiative transfer
method on \citet{Laursen2}.  In this method, \lya{} photons are
isotropically initialized at the sphere center for the uniform case,
and in radial shells for the non-uniform cases.  They have a relative
frequency
\begin{equation}
  x_{\rm ph} = x_{\rm Ly\alpha} - \mathbf{v}_{\rm H} \cdot
  \hat{\mathbf{n}}_{\rm ph},
\end{equation}
where $\mathbf{v}_{\rm H}$ and $\hat{\mathbf{n}}_{\rm ph}$ are the
bulk velocity of the gas in units of the sound speed and the photon
propagation direction, respectively.  We then transport each photon
according to the following prescription.  The photon travels along a
direction $\hat{\mathbf{n}}_{\rm ph}$ for a distance $r$ that
corresponds to a uniformly distributed random optical depth $\tau =
\int N_{\rm HI} \sigma(x) r \, dr$, where $\sigma(x)$ is the \lya{}
cross-section at the relative frequency $x$ and $N_{\rm HI}$ is the
neutral hydrogen column density.  During an interaction, the photon
scatters off a neutral hydrogen atom, causing a frequency shift
$\Delta x = -u_\parallel + \hat{\mathbf{n}}_{\rm ph} \cdot
\mathbf{u}$.  Here $\mathbf{u}$ is the relative velocity between the
gas and photon, and $u_\parallel$ is the component parallel to
$\hat{\mathbf{n}}_{\rm ph}$.  After the photon is scattered, the
probability of a change in propagation direction $\theta$ is given by
the phase function
\begin{equation}
  W(\theta) =
  \begin{cases}
    1/2, & \text{(core;}\ 2P_{1/2}) \\
    (7/16)[1 + (3/7)\cos^2\theta)], & \text{(core;}\ 2P_{3/2}) \\
    (3/8)(1 + \cos^2\theta), & \text{(wing)}
  \end{cases}
\end{equation}
that is derived from a dipole approximation of the interaction, and
in the profile wings, the scattering behaves like a classical system
producing a dipole distribution \citep{Hamilton40, Stenflo80, Laursen2}.

\section{Results}
\label{sec:3}
\begin{figure}
  \includegraphics[width=\columnwidth]{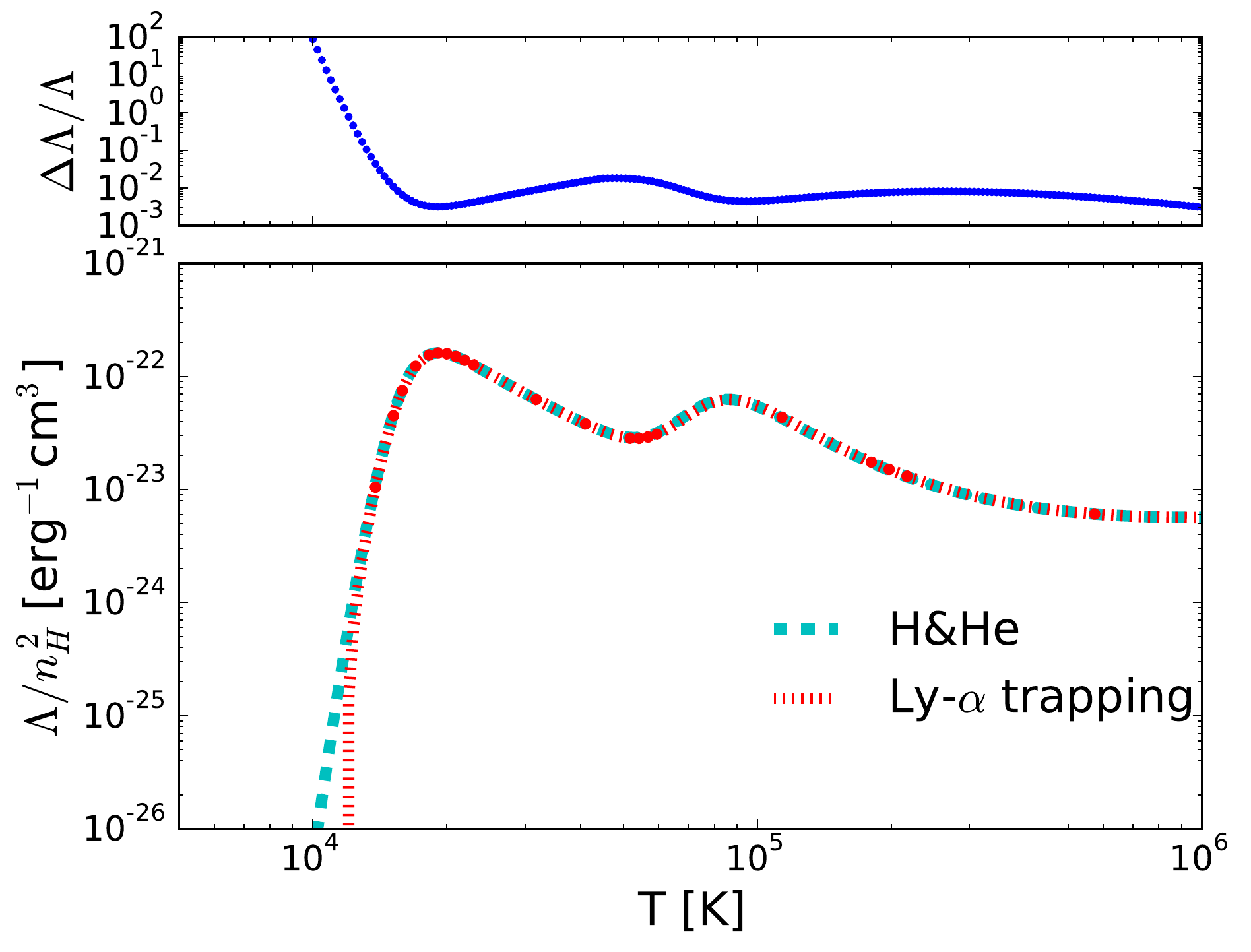}
  \caption{Comparison of overall cooling rates in the optically thin
    and thick approximations.  Top panel: The fractional difference in
    cooling rates between the two cases.  Bottom panel: Dependence of
    the primordial cooling rate per $n^2$ on temperature in the
    optically-thin limit (blue dashed) and while considering \lya{}
    trapping (red dotted).  Significant differences only exist at $T <
    20,000\unit{K}$ because gas is optically thick to \lya{}
    radiation.}
  \label{fig:sahacooling}
\end{figure}

\subsection{Radiative cooling with \lya{} trapping} 
\label{sec:cooling-results}

The massive seed BH mass is largely influenced by the mass accretion
rates into the central gas cloud and any fragmentation that might
occur during its catastrophic collapse, prompted by the radiative
cooling of a primordial gas.  The virial temperature of the candidate
halo that hosts massive BH seed formation is $\gtrsim 8000 \unit{K}$,
which corresponds to a virial mass $M_{\rm vir} \simeq 10^8~\Ms{}
[(1+z)/10]^{-3/2}$.  For such a contraction to proceed the cooling
timescale 
\begin{equation}
  \label{eq:collapse}
  t_{\rm cool} \simeq \frac{1}{n_{\rm e}^2 \Lambda(T)} \frac{3 \rho
    kT}{2 \mu m_{\rm p}}
\end{equation}
must be shorter than or similar to the dynamical time $t_{\rm dyn} =
(G\rho)^{-1/2}$ \citep{White78}.  Here $\Lambda$ is the cooling
function, and $n_e$ and $\rho$ are the electron number density and gas
density, respectively.  Starting at temperatures $T \sim 10^4
\unit{K}$, hydrogen becomes partially ionized, eventually reaching
near complete ionization at $T \sim 1.5 \times 10^4 \unit{K}$.  Thus in
these halos with virial temperatures near this limit, the assumption
that the primordial gas is either completely ionized ($\mu = 0.6$) or
neutral ($\mu = 1.22$), in addition to $n_{\rm e}$, could be
inaccurate during the collapse and should be tracked.

At low densities $n \lesssim 100~\cubecm$, the use of the optically thin
cooling rates is valid.  However when \lya{} radiation from
collisional and recombination processes is extremely attenuated at
higher densities, the cooling function $\Lambda$ should decrease as
thermal energy cannot be effectively radiated out of the system
anymore.  When these cooling channels are blocked, a primordial gas
can still radiatively cool through the two-photon process.

Figure \ref{fig:sahacooling} compares the cooling function of atomic
metal-free gas in the optically-thin regime and when the gas is
optically-thick to \lya{} radiation.  The trapped \lya{} radiation
reduces the cooling rates at $T \lesssim 2 \times 10^4 \unit{K}$,
which could result in higher temperatures as the primordial gas cloud
collapses.  Above this temperature, \lya{} trapping and the associated
resonance scattering does not occur because spontaneous emission in
hydrogen dominates, and furthermore helium de-excitation cooling
becomes important at these higher temperatures.  Nevertheless, we next
investigate this effect further in our \lya{} radiative transfer
calculations as the system is dynamically collapsing, checking how the
thermodynamic properties change during this event.


\subsection{Cosmological Halo Collapse: A Basis for \lya{} Transfer}
\label{sec:cosmo}

\begin{figure} 
  \includegraphics[clip, width=\columnwidth]
    {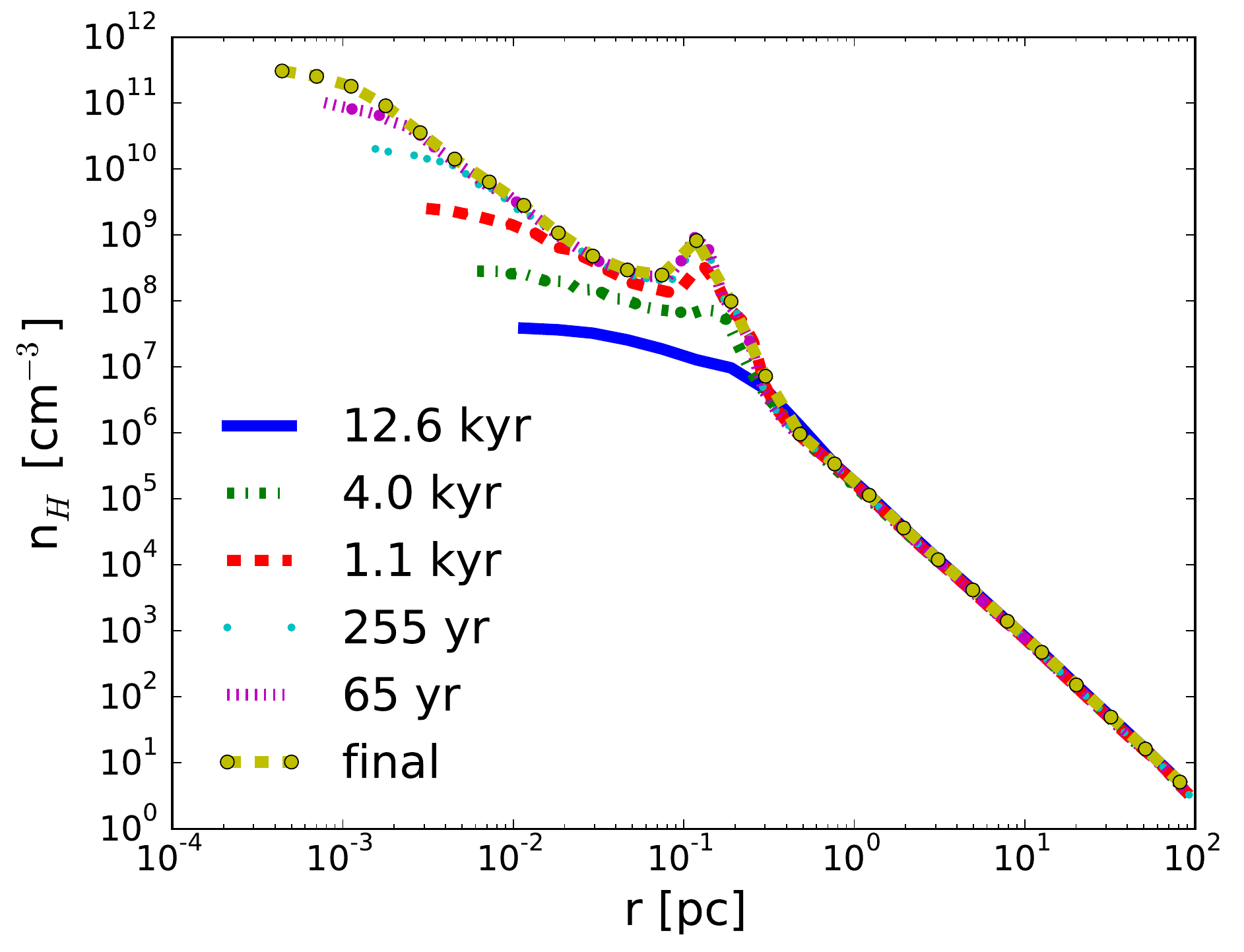}
  \includegraphics[clip, width=\columnwidth]
    {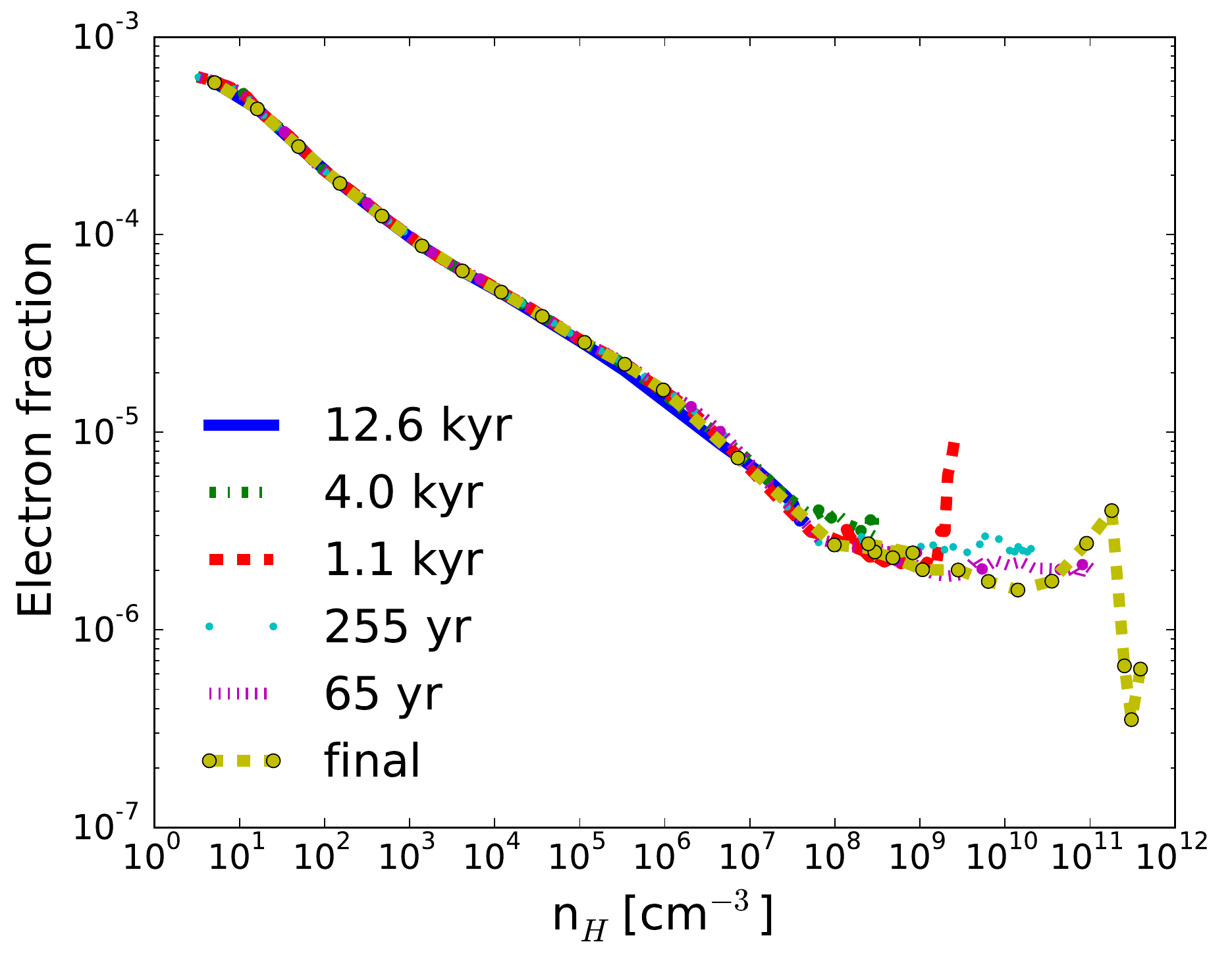}
    \caption{Radially-averaged profiles of gas number density (top)
      and profiles of electron fraction with respect to number density
      (bottom) at the final output when the collapse reaches $n_{\rm
        H} = 3 \times 10^{11} \cubecm$ and 65, 255, 1,100, 4,000, and
      12,600 years before this time.  The halo density follows a
      $r^{-2.2}$ power law, appropriate for an isothermal collapse.
      The bump at 0.1~pc corresponds to lesser overdensity that has
      fragmented from the main collapsing cloud.  The electron
      fraction drops with density as free electrons are consumed by
      recombinations.}
  \label{fig:profile}
\end{figure} 

We utilize a collapsing halo from a cosmological simulation as the
basis for the Monte Carlo radiation transfer calculations, providing a
more realistic environment for the propagation medium of the \lya{}
photons.  This halo is the most massive in the simulation domain with
a total mass $M_{\rm tot} = 5.85 \times 10^7~\Ms$ and a virial radius
$r_{\rm vir} = 782 \unit{pc}$ when it catastrophically collapses at $z
= 14.664$.  This halo mass corresponds to a virial temperature $T_{\rm
  vir} = 1.17 \times 10^4 \unit{K}$, which is typical of a metal-free
atomic cooling halo that cools and collapses for the first time.  The
halo does not experience any major mergers for the last 100~Myr
of the simulation.

Figure \ref{fig:profile} shows the evolution of radially-averaged
profiles of the gas number density $n_{\rm H}$ and the average
electron fraction at a given $n_{\rm H}$ value.  The first density
profile depicts the system when the maximum $n_{\rm H} \simeq
10^{7.5}~\cubecm$ (AMR level 15), and then the profiles are shown as
the maximum density increases by $\sim$1~dex, finally reaching a
maximum $n_{\rm H} \simeq 3 \times 10^{11}~\cubecm$ (AMR level 20).
The density profile generally exhibits a power law $\rho \propto
r^{-2.2}$ from the virial radius to $\sim 10^{-3}\unit{pc}$.  This
feature is typical of an isothermal collapse, which happens in this
case at $T \simeq 8000\unit{K}$, where the gas cooling is limited to
atomic processes in the presence of a strong LW radiation field
$J_{21} = 10^5$.  With the isothermal density profile, the inner 1~pc
is gravitationally unstable, and its Jeans mass is $\sim 10^5~\Ms$,
similar to previous works \citep[e.g.][]{Wise, Regan, Shang,
  Becerra15}.  One exception to the centrally concentrated,
spherically symmetric collapse is a clump that fragments $\sim
0.1\unit{pc}$ from the densest point, seen as a bump in the density
profile, which initially fragments about 5~kyr before the collapse.
The electron fraction in the lower panel of Figure \ref{fig:profile}
shows that the free electron fraction decreases with density
(i.e. radius) as the recombination rate increases with $n^2$,
eventually saturating at $2 \times 10^{-6}$.  The electron fraction
will play an important role in determining the \lya{} emissivity as it
is directly related to the electron number density (Equations
\ref{eqn:rec} and \ref{eqn:dcol}).


\subsection{Monte Carlo Radiation Transfer}
\label{sec:mcrt}

\begin{figure}
  \includegraphics[width=\columnwidth]{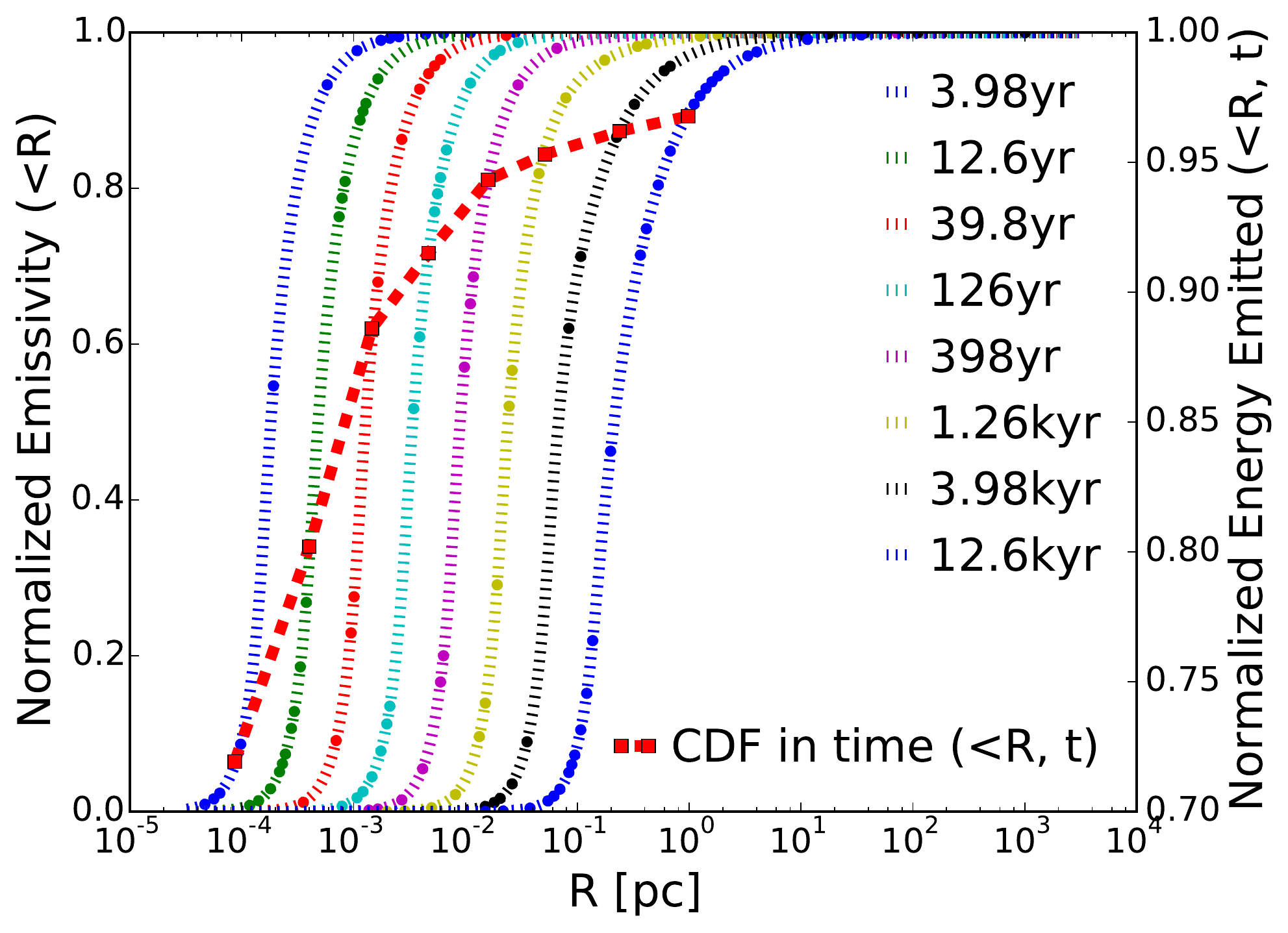}
  \caption{{\it Dotted lines and left axis:} Normalized \lya{}
    emissivity cumulative profiles at eight times (see legend) before
    the final simulation output.  {\it Dashed line and right axis:}
    Cumulative amount of \lya{} radiation emitted from the time
    indicated by the intersecting dotted line to 1~Myr before the
    final output time, i.e. 96\% of all \lya{} radiation is emitted
    within 4~kyr of the final collapse.}

  \label{fig:photongeneration}
\end{figure}

\begin{figure*}
  \includegraphics[width=\textwidth]{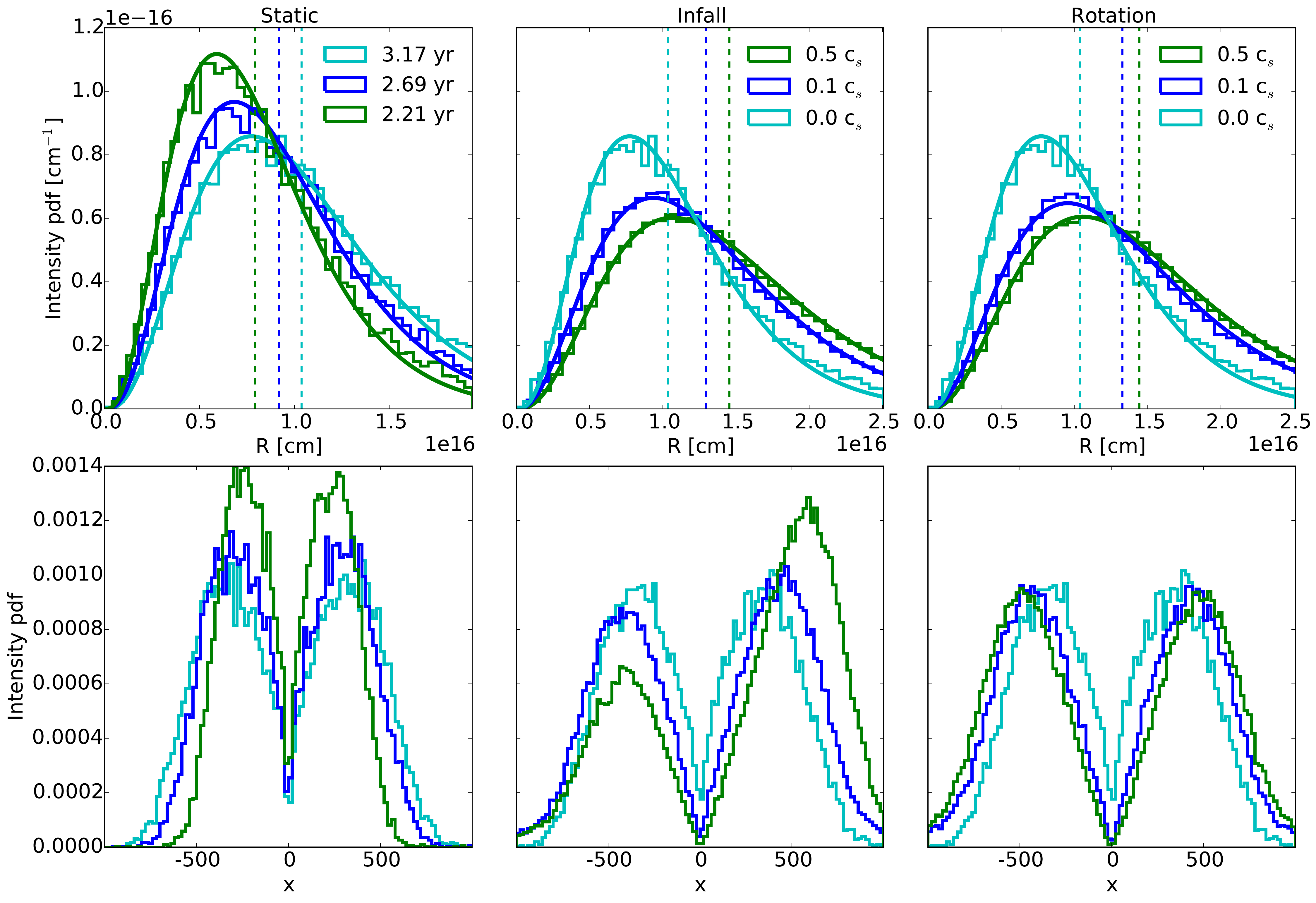}
  \caption{Ly$\alpha$ radiation transfer calculations in the uniform
    density case with a hydrogen number density $n_{\rm H} =
    10^8~\cubecm$. {\it Top row:} \lya{} radiation intensity as a
    function of radius for the static case (left), infall case
    (middle), and rotation case (right), where the vertical lines show
    the intensity-weighted means.  The histograms depict the results
    from the radiation transport calculation, whereas the smooth curve
    is a fit to the distribution.  The static case shows the radiation
    propagating outwards, slower than the speed of light due to
    scattering, at three different times.  The infall and rotation
    cases, shown at $t = 3.17 \unit{yr}$ ($ct = 3 \times 10^{16}
    \unit{cm}$), demonstrate that the bulk motion of the gas allows
    the radiation to propagate farther as the photons experience a
    greater Doppler shift when they are re-emitted.  {\it Bottom row:}
    The normalized spectra of \lya{} radiation for the same cases
    shown in the top row.  In the static case, the \lya{} photons
    shift away from the line center as time progresses, and the infall
    and rotation cases show the increased Doppler shifts as the gas
    bulk motion increases.}
  \label{fig:uniform}
\end{figure*} 

Before invoking a radiation transport calculation, we first calculate
the \lya{} emissivity, using Equations (\ref{eqn:rec}) and
(\ref{eqn:dcol}), in the collapsing halo at eight different snapshots
during the event.  The bulk of the emission occurs in the central
regions, as expected, and we show the cumulative \lya{} luminosity,
\begin{equation}
\label{eqn:gen}
L_{\rm Lya}(<r, t) = \int_0^r \left[ \eta^{\rm rec}(t) + \eta^{\rm
    dcol}(t) \right] dV, 
\end{equation}
as function of radius in Figure \ref{fig:photongeneration} with the
sphere centered on the densest point.  One can see that as the inner
region collapses, the source of \lya{} emission shrinks as the density
increases.  When the central object becomes gravitationally unstable
12.6~kyr before the final time, 90\% (50\%) of the emission comes from
the central 1.0 (0.2) pc.  This radius decreases gradually with time
until a sphere of radius $10^{-3}$ generates 99\% of the \lya{}
radiation at the final time.  

As the halo is dynamically collapsing, we can calculate the total
\lya{} energy being emitted throughout its collapse by numerically
integrating Equation (\ref{eqn:gen}) from 1~Myr before the final
simulation time to the times shown in Figure
\ref{fig:photongeneration}.  The red dashed line shows this value, and
the differences between adjacent points equal the percentage of total
\lya{} radiation generated between these two times.  For instance, at
4~kyr before the final time, 96\% of all \lya{} radiation during the
collapse is generated after this time, with most of the photons
originating within a radius 0.1~pc.  This fractional energy decreases
with time until 72\% of the \lya{} radiation originates only 4~yr
before the final collapse.

Both the location and timing of the \lya{} radiation will aid us in
constructing Monte Carlo calculations with the appropriate length and
temporal scales.  These simulations will explore physical scenarios
that gradually increase the realism of the environment through which
the \lya{} photons propagate.  First we will inspect the uniform
density case, then the time-independent isotropic case, and finally a
collapsing time-dependent isotropic case.  The last and most realistic
case is used to calculate the effective equation of state, which is an
essential ingredient when determining the thermodynamic behavior, and
thus possible fragmentation, of the collapsing system.

\subsubsection{Uniform density case}

\begin{table}
  \caption{Fitting parameters for the radiation distribution in the
    uniform density case}
  \centering
  \label{tab:uniform}
  \begin{tabular*}{0.99\columnwidth}{@{\extracolsep{\fill}} ccccc }
    \hline 
    Case & $\log (n_{\rm H} / \cubecm)$ & $a$ & $b_0$ & $b_1$\\
    \hline
    \multirow{4}{*}{Static} & 6 & 3.46 & $1.28 \times 10^{10}$ & 0.732\\
    & 7 & 3.48 & $6.67 \times 10^9$ & 0.738\\
    & 8 & 3.48 & $2.42 \times 10^9$ & 0.730\\
    & 9 & 3.50 & $4.15 \times 10^9$ & 0.731\\
    \hline
    \multirow{4}{*}{Infall} & 6 & 2.80 & $2.04 \times 10^{10}$ & 0.747\\
    & 7 & 2.85 & $8.86 \times 10^9$ & 0.756\\
    & 8 & 2.61 & $3.40 \times 10^9$ & 0.750\\
    & 9 & 2.60 & $1.24 \times 10^9$ & 0.768\\
    \hline
    \multirow{4}{*}{Rotation} & 6 & 2.77 & $1.90 \times 10^{10}$ & 0.749\\
    & 7 & 2.72 & $1.02 \times 10^{10}$ & 0.752\\
    & 8 & 2.77 & $2.98 \times 10^9$ & 0.753\\
    & 9 & 2.78 & $1.65 \times 10^9$ & 0.757\\
    \hline
  \end{tabular*}
  \vspace{0.5em}
  \parbox[t]{\columnwidth}{Notes: The parameters apply to Equation
    (\ref{eqn:gamma-uniform}). The static case has zero bulk
    velocity.  The parameters for the infall and
    rotation case are shown only for the $0.5 c_{\rm s}$ cases.}
\end{table}


The most fundamental case to inspect in a \lya{} transfer calculation
is a gas parcel with uniform density and temperature.  Here we monitor
how the radiation propagates from a single impulse originating from
a point source at $r=0$.  We execute a series of simulations in
spherical symmetry with a uniform temperature of 8000~K, which is
similar to the temperatures in the atomic cooling halo presented in
Section \ref{sec:cosmo}, and four different hydrogen number densities
$\log(n_{\rm H}/\cubecm) = (6, 7, 8, 9)$.  The top row of Figure
\ref{fig:uniform} shows the radial behavior of the radiation energy
distribution in the $n_{\rm H} = 10^8~\cubecm$ case for the static
(left panel), infall (middle panel), and rotation (right panel) cases.

The static case, which is shown at three times, $t = (2.21, 2.69,
3.17) \unit{yr}$ with the last time corresponding to $t = 10^8
\unit{s}$ and a light travel time $ct = 3 \times 10^{18} \unit{cm}$,
have the radial distributions that are well fit with Gamma
distributions, valid for the entirety of the simulation time $t < 10^8
\unit{s}$,
\begin{equation} 
  \label{eqn:gamma-uniform}
  p(r,t) = \frac{r^{a-1} \, e^{-r/b(t)}}{[b(t)]^{a} \, \Gamma(a)}.
\end{equation} 
Here $a$ is a constant and controls the distribution width (i.e. the
shape parameter), and $b(t) = b_0 t^{b_1}$ varies with time (i.e. the
rate parameter) and controls the length of the tail at larger radii.
$\Gamma(x)$ is the complete Gamma function, and $t$ is in units of
seconds.  We do not consider the distribution beyond a light travel
time $r = ct$.  These parameters are given in Table \ref{tab:uniform}.
For such a distribution, the maximum value occurs at $(a-1)b$; the
average value is $ab$; the skewness is $2/\sqrt{a}$.  Taking the
$n_{\rm H} = 10^8~\cubecm$ case as an example, we have $ab = 0.281 \,
t^{-0.272} \times (ct)$.  Compared to the optically thin case ($ab =
ct$), the \lya{} radiation is diluted by a factor of 0.281, and its
propagation slows as time progresses, as indicated by the negative
exponent.  This behavior is apparent in the top-left panel of Figure
\ref{fig:uniform}, where the distribution migrates to larger radii
with its tail becoming longer.  Looking at other densities, the shape
parameter $a$ is basically unchanged, which is analogous to having a
resonant scattering shell with a constant relative thickness.  The
bottom row of Figure \ref{fig:uniform} shows the relative frequencies
of the photons.  In the static case, the spectrum is symmetric around
the line center ($x=0$), which is expected, and obeys the
\citet{Neufeld} profile.  The width of the lines depend on the optical
depth of the system and the time elapsed.  At early times, the photons
are nearest to the line center, and they Doppler shift away from the
center as they resonate in the neutral hydrogen medium.

\begin{figure*}
  \includegraphics[width=\textwidth]{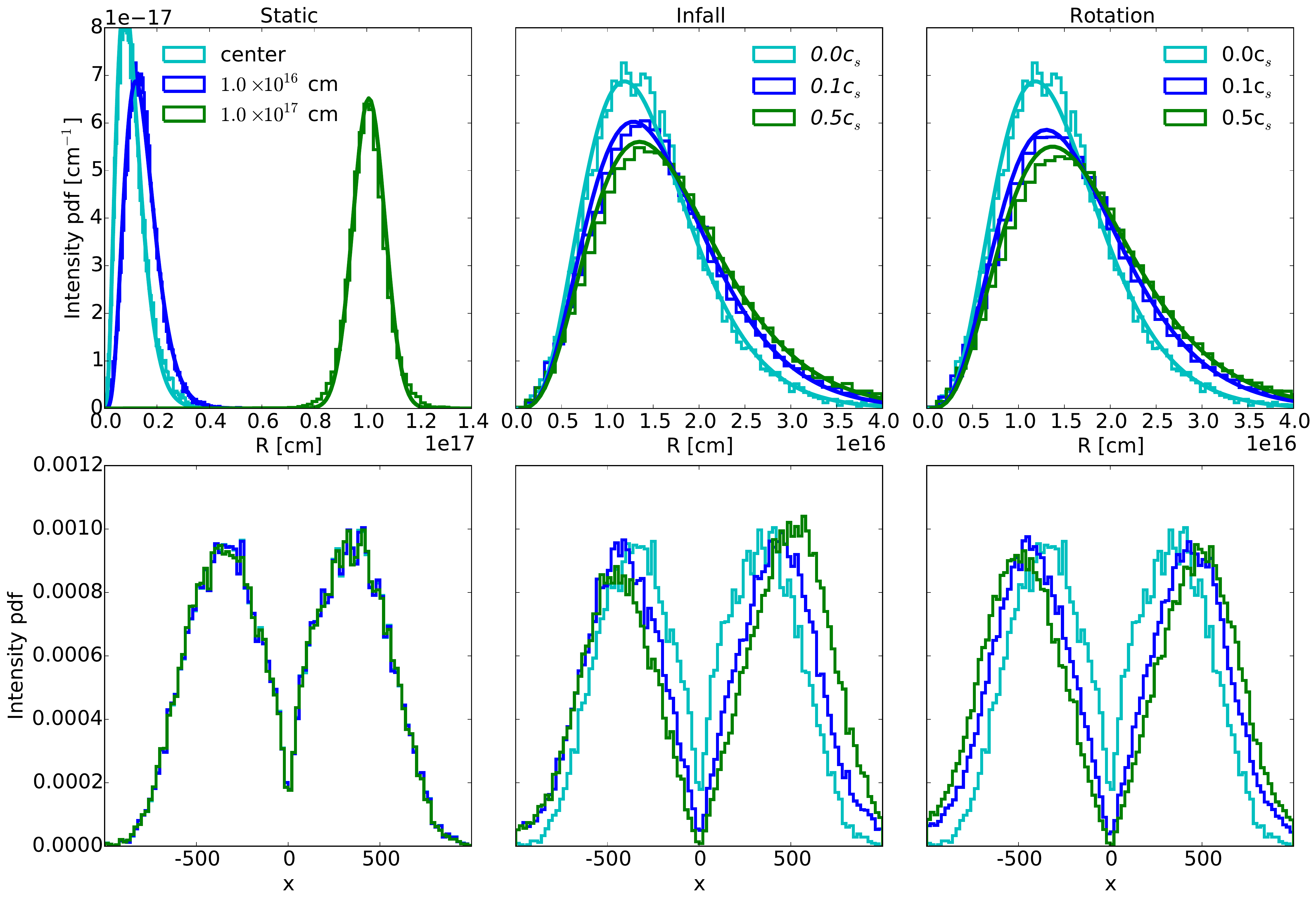}
  \caption{Same as Figure \protect\ref{fig:uniform} but for the
    time-independent isotropic case that propagates \lya{} radiation
    through a spherically symmetric halo with its quantities taken
    from a cosmological simulation.  The static case (left) shows the
    radiation distribution and spectra at $t = 10^8 \unit{s}$ when the
    photons are generated at the halo center and in two shells with
    radii $r = 10^{16} \unit{cm}$ and $10^{17} \unit{cm}$.  The
    radiation preferentially propagates outward because of the
    density gradient.  The infall (center) and rotation (right) cases
    are shown at the same time with the photons generated in a shell
    of radius $r = 10^{16} \unit{s}$ for speeds $v/c_{\rm s} = (0.1,
    0.5)$.  Their distributions and spectra show similar behavior as
    the uniform density case with the photons being Doppler shifted as
    the velocity increases, resulting in a wider radiation distribution.}
  \label{fig:isotropic}
\end{figure*}

Next we inspect the radial distribution of \lya{} radiation and its
spectra in the infall and rotation cases, which are shown in the
middle and right columns of Figure \ref{fig:uniform}.  Comparing the
spectra of the infall cases with $v_{\rm r}/c_{\rm s} = (0.1, 0.5)$
and the static case, we see that the photons are blue-shifted farther
away from the line center, which occurs when the infalling gas
re-emits the \lya{} photons whose relative velocity causes an increase
in frequency.  Because of the enhanced Doppler shift, the photons
scatter less because of the decreased optical depth away from the line
center, allowing for the \lya{} radiation to propagate farther away
from the sphere center, which is seen in a broader radial profile.  In
the rotation case with $v_\theta/c_{\rm s} = (0.1, 0.5)$, the photons
are symmetrically shifted into the wings of the line, which extends
the radiation distribution similar to the infall case.  These
distributions are still nicely fit with a Gamma distribution (Equation
\ref{eqn:gamma-uniform}) at various number densities and bulk
velocities, and we show the fitting parameters in Table
\ref{tab:uniform} alongside the static case.  Both infall and rotation
cases have larger $b$ parameters, indicating that the radiation is
less trapped in the gas.

\begin{figure*}
  \includegraphics[width=\textwidth]{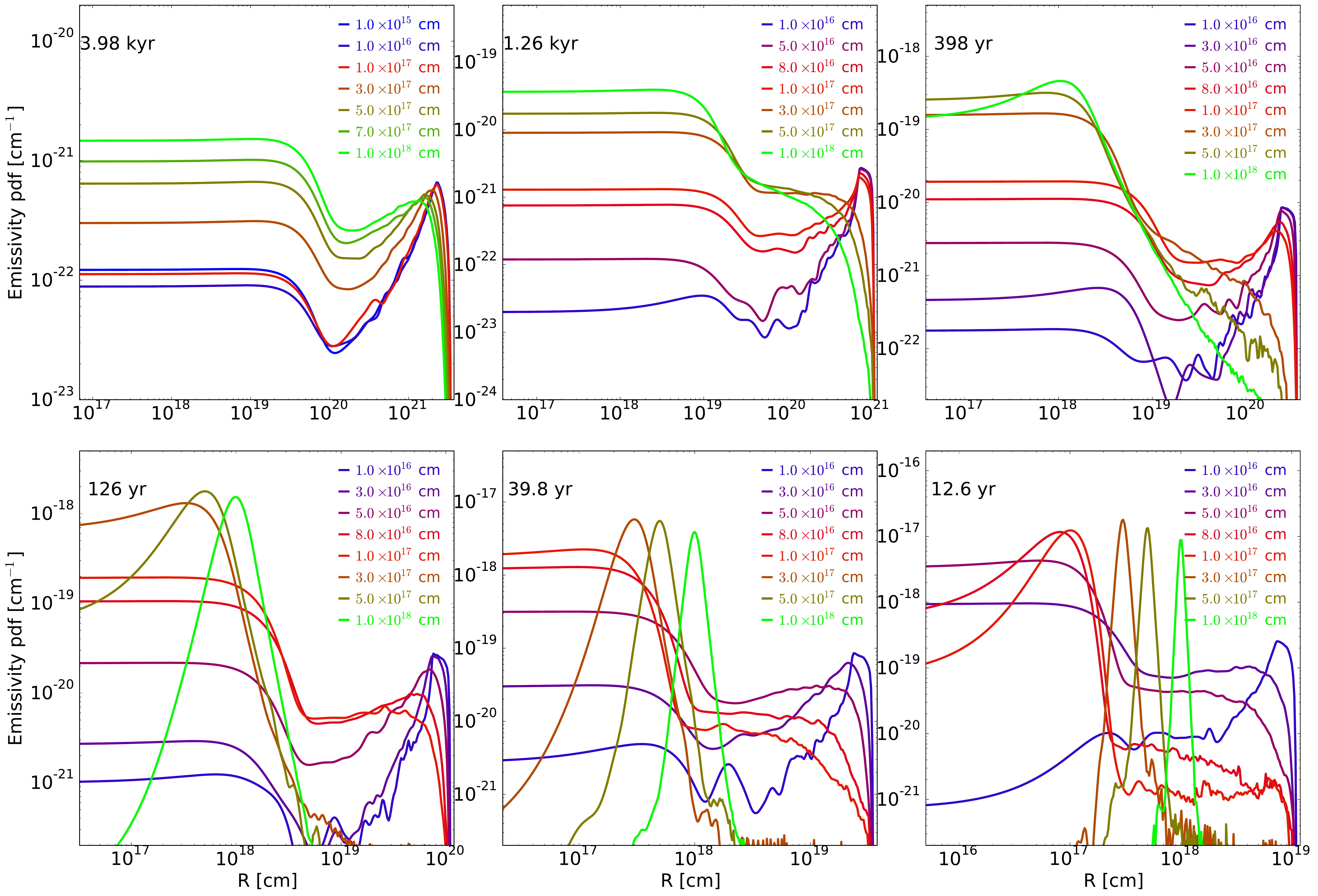}
  \caption{Radial distributions of \lya{} emissivity sourced from a
    subset of individual shells in the time-dependent isotropic case
    at six different times before the final collapse.  The lines are
    colored by their shell radius and track the propagation from that
    initial radius to the final state.  At early times, there is
    enough time for the emissivities to reach an equilibrium in the
    core while some fraction escape into the outer regions of the
    halo.  At later times, the distributions from the larger shells do
    not have time to propagate far from their origin, while the
    smaller shells contribute the most to the \lya{} radiation field
    in the inner core with radius 0.1~pc.}
  \label{fig:cumulative}
\end{figure*}


Because a single impulse of radiation sourced these radial
distributions, we can increase the realism of the calculation by
integrating these radiation profiles with respect to time in the range
$t = 0 \rightarrow \mathcal{T}$, which representing the center
constantly emitting photons.  Integrating Equation
(\ref{eqn:gamma-uniform}), we find that the radiation profile
transforms into
\begin{equation} 
  \label{eqn:gamma-static}
  P(r,\mathcal{T}) = \int_0^{\mathcal{T}} p(r, t) \, dt =
  \frac{r^{1/(b_1-1)} \, \Gamma(a-1/b_1,
    (r/b_0) \, \mathcal{T}^{-b_1})} 
  {\mathcal{T} \, b_1 \, b_0^{1/b_1} \, \Gamma(a)},
\end{equation} 
where $\Gamma(x,y)$ is the incomplete Gamma function.  The resulting
distribution has an intensity-averaged radius $\mathbb{E}(r) =
ab/(1+b_1)$ that is proportional to the distribution coming from a
radiation impulse.  In this case, the scattering and trapping of
\lya{} radiation and its diminished radial propagation declines with
time as $t^{-\alpha}$, where $\alpha = 2.4-2.7$, in the static case.

\subsubsection{Time-independent isotropic case} 

Now that we have established the behavior of \lya{} radiation
transport in a uniform density and temperature case, we turn our
attention to the time-independent isotropic case.  Here we consider a
spherically symmetric system, taking the radial profiles of density
and temperature, along with the average electron fraction as a
function of density, from the atomic cooling halo in the cosmological
simulation (Section \ref{sec:cosmo}), 4.0~kyr before the final output
when the maximum number density $n_{\rm H} = 3 \times 10^8~\cubecm$.
In this case, we consider three cases of \lya{} radiation generation:
from the center of the halo and from two concentric shells with radii
$r = 10^{16} \unit{cm}$ and $10^{17} \unit{cm}$.  We do not utilize
the velocity information from the simulation but consider the same
velocity setups as the uniform density case: static, infalling, and
rotation, where the latter two configurations have $v/c_{\rm s} =
(0.1, 0.5)$.  We allow these photons to propagate according to the
scattering radiative transfer equation into the spherically symmetric
halo.

\begin{table}
  \caption{Fitting parameters for the radiation distribution in the
    static isotropic case}
  \centering
  \label{tab:isotropic}
  \begin{tabular*}{0.99\columnwidth}{@{\extracolsep{\fill}} cccccc }
    \hline
    Case & $r$ [cm] & $a_0$ & $a_1$ & $b_0$ & $b_1$\\
    \hline
    \multirow{3}{*}{Static} & 0 & $5.99 \times 10^7$ & 0.801 & $6.90 \times 10^{-7}$ & 2.448\\
    & $10^{16}$ & $7.87 \times 10^8$ & 0.916 & $1.24 \times 10^{-7}$ & 2.525\\
    & $10^{17}$ & $6.48 \times 10^9$ & 0.996 & $1.21 \times 10^{-3}$ & 2.162\\
    \hline
    \multirow{3}{*}{Infall} & 0 & $3.46 \times 10^2$ & 0.259 & $2.26 \times 10^6$ & 1.180\\
    & $10^{16}$ & $3.98 \times 10^2$ & 0.265 & $1.71 \times 10^{6}$ & 1.193\\
    & $10^{17}$ & $7.48 \times 10^2$ & 0.291 & $3.44 \times 10^{6}$ & 1.162\\
    \hline
    \multirow{3}{*}{Rotation} & 0 & $2.98 \times 10^2$ & 0.248 & $1.90 \times 10^{6}$ & 1.156\\
    & $10^{16}$ & $3.82 \times 10^2$ & 0.280 & $2.02 \times 10^{6}$ & 1.176\\
    & $10^{17}$ & $8.44 \times 10^2$ & 0.301 & $2.98 \times 10^{6}$ & 1.142\\
    \hline
  \end{tabular*}
  \vspace{0.5em}
  \parbox[t]{\columnwidth}{Notes: The parameters apply to Equation
    (\ref{eqn:gamma-uniform}) but with $a(t) = a_0 \, t^{-a_1}$. For
    $r = 0$, the \lya{} radiation originates at the halo center, and
    for the non-zero radii, it originates from shells of those radii.
    The static case has zero bulk velocity.  The parameters for the
    infall and rotation case are shown only for the $0.5 c_{\rm s}$
    cases.  The associated fits are accurate for $t < 10^9 \unit{s}$.}
\end{table}


Figure \ref{fig:isotropic} shows the resulting radiation radial
distribution (top panels) and spectra (bottom panels) at a time $t =
10^8 \unit{s} = 3.17 \unit{yr}$, corresponding to a light travel time
$ct = 3 \times 10^{18} \unit{cm}$.  Focusing first on the static case
(left column), the \lya{} radiation propagates away from the center
with a maximum at $8 \times 10^{15} \unit{cm}$, while the photons from
the radiating shells at $r = 10^{16} \unit{cm}$ and $10^{17}
\unit{cm}$ preferentially migrates outward because of the density
gradient.  These distributions are again well fit with a Gamma
distribution (Equation \ref{eqn:gamma-uniform}), similar to the
uniform density case but with $a(t) = a_0 t^{-a_1}$ instead of being a
constant.  The fitting parameters for the three cases are given in
Table \ref{tab:isotropic}.  The preference toward outward propagation
can be quantified by inspecting the skewness ($2/\sqrt{a}$) of these
distributions, which are in the range 0.2--0.4.  The spectra for the
central and shell sources have similar spectra, as expected, because
they are shown at the same integration time and the velocities are the
same.

Both the radial distribution of \lya{} radiation and the spectra of
the infall and rotation cases behave similarly to their counterparts
in the uniform density case.  The middle and right columns of Figure
\ref{fig:isotropic} show these respective cases at a time $t = 10^8
\unit{s}$ with the photons being generated in a shell of radius $r =
10^{16} \unit{cm}$.  We also consider the cases where photons are
generated in the center and a larger shell of radius $10^{17}
\unit{cm}$, whose fitting parameters are shown in Table
\ref{tab:isotropic} but not shown in the Figure.  The relative
velocities of the gas cause a Doppler shift, allowing the photon
frequencies to migrate away from the line center, with a tendency
toward a blueshift in the infall case and symmetric shifts in the
rotation case.  This effect increases their mean free path, extending
the radial profiles.  As the photons propagate outward into the more
diffuse regions (recall $\rho \propto r^{-2.2}$) of the halo, they will
scatter less frequently, eventually free streaming away from the halo
center.

\subsubsection{Time-dependent isotropic case} 

We now consider the case where the halo is dynamically collapsing,
whereas previously we restricted the integration times to $10^8
\unit{s}$ that is comparable to the light-crossing time of the inner
parsec.  This time-dependent calculation is similar to the
time-independent calculation; however we utilize six outputs from the
cosmological simulation that are evenly log-spaced in time, starting
at 3.98~kyr before the collapse.  The density profiles are
approximately isothermal with $\rho \propto r^{-2.2}$ at all times
with the maximum density increasing from $3 \times 10^8~\cubecm$ to $3
\times 10^{11}~\cubecm$ during this time.  At each time, 20 shells
radiate \lya{} photons, which are equally log-spaced in radius ranging
from $10^{14} \unit{cm}$ to $10^{19}\unit{cm}$.  The output times
$t_i$ are given in Table \ref{tab:dynamic}.  The largest shell
encloses nearly all of the \lya{} radiation that was depicted in
Figure \ref{fig:photongeneration}.  The major improvement upon the
previous cases is that we integrate over the resulting radiation
distribution from each shell to compute a cumulative radiation
distribution for the entire halo.

Starting at the earliest time, the shells radiate for a time equal to
the duration between outputs (i.e. $t_i - t_{i+1}$).  We track the
radial distribution of the photons from each shell, where Figure
\ref{fig:cumulative} shows a subset of the 20 shells, allowing us to
inspect the propagation behavior from each radiation origin.  At the
earliest time (3.98~kyr), the radial distributions from each shell
have similar shapes.  They have plateaus at small radii, which have
reached an equilibrium between emission and scattering out of the
center.  The local maxima at $r \simeq 1-2 \times 10^{21} \unit{cm}$
represent the photons that have escaped the inner regions by
scattering many times and driving the frequency into the wings of the
spectrum, and they are freely streaming outward through the diffuse
outer regions.  As the collapse progressively accelerates, the
dynamical time decreases, giving less time for the photons to
propagate through the pre-galactic medium.  This behavior can be seen
through the steadily decreasing radiation distribution at large radii
at $t = 1260$ and $398 \unit{yr}$ for the largest shells.  Eventually
in the latter times, the distributions for the largest shells
transform into Gamma distributions as the integration times shorten to
$\sim 10^8 \unit{s}$.  However for the radiation originating from
intermediate radii (e.g. $r = 3 \times 10^{16} \unit{cm}$ and $5
\times 10^{16} \unit{cm}$ at $t = 12.6\unit{yr}$), the \lya{} photon
distribution still have a plateau at small radii.

\subsection{Effective equation of state with \lya{} scattering}

\begin{table}
  \caption{Output times and range of radii of the radiating shells in
    the equation of state calculation}
  \centering
  \label{tab:dynamic}
  \begin{tabular*}{0.99\columnwidth}{@{\extracolsep{\fill}} c | cc }
    \hline 
    Case $i$ & Time $t_i$ [yr] & Shell radii range [$10^{16}$ cm]\\
    \hline 
    0 & 3,980 & 10 -- 100\\ 
    1 & 1,260 & 5 -- 100\\
    2 & 398   & 3 -- 100\\
    3 & 126   & 2 -- 100\\
    4 & 39.8  & 1 -- 100\\
    5 & 12.6  & 1 -- 100\\
    \hline
  \end{tabular*}
\end{table}


The radiation distributions from individual shells informs us how the
radiation transports given an origin, but at some given time, the
overall \lya{} emissivity distribution is the key quantity in
determining the coupling between the \lya{} photons and the neutral
medium.  Ultimately, we can compare the radiation distribution from
the transport calculation to the optically-thin (free streaming) case
to calculate the reduction in the radiative cooling rate from
collisional excitations and ionizations.  

\begin{figure}
  \includegraphics[width=\columnwidth]{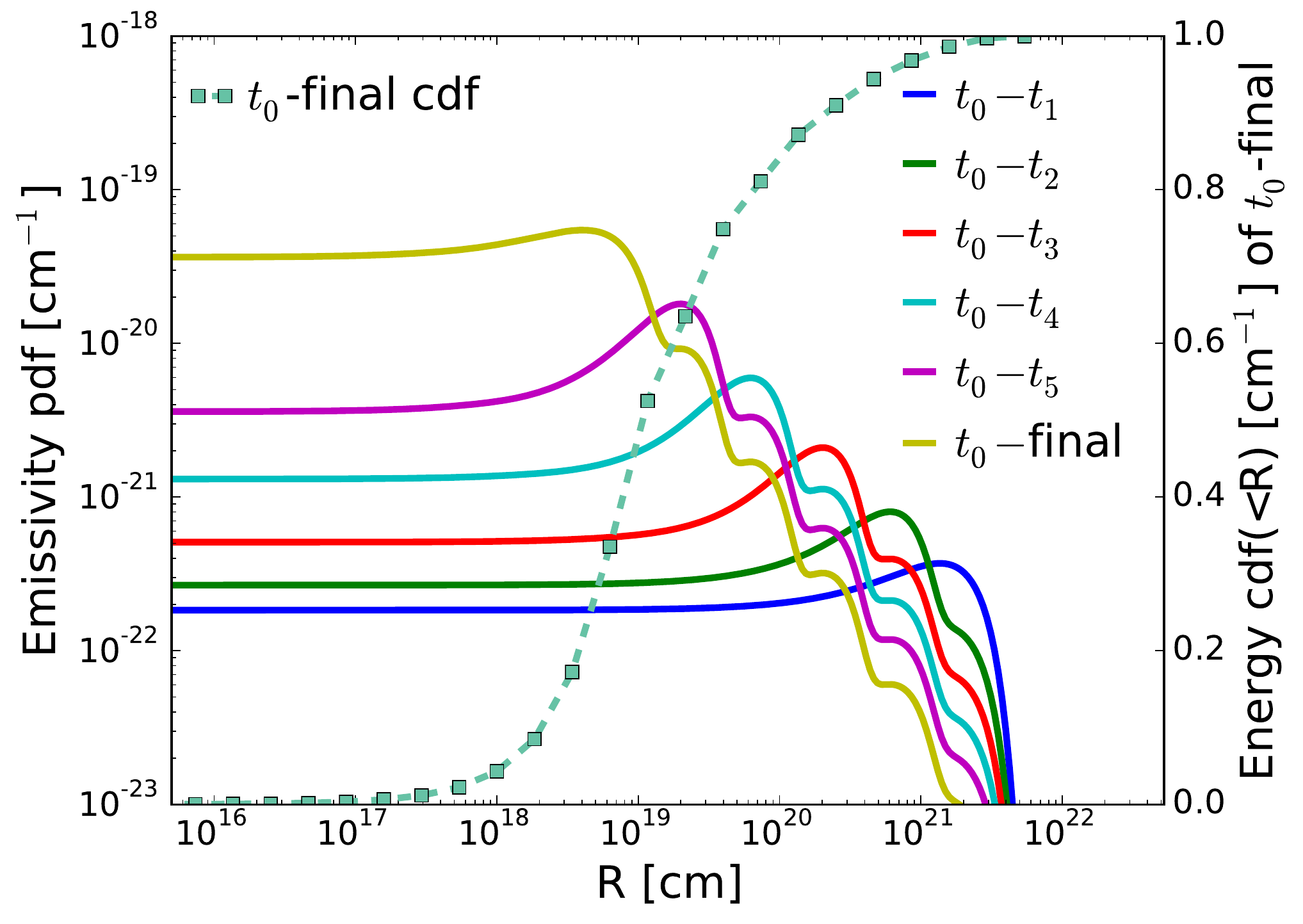}
  \caption{The total \lya{} emissivity integrated (Equation
    \protect\ref{eqn:total_eta}) from all of the shells shown in
    Figure \protect\ref{fig:cumulative} and all times from $t_0$ to
    $t_i$.  The yellow (top) solid line shows the \lya{} emissivity at
    the final time of the simulation.  The dashed line with square
    points depicts the cumulative \lya{} emissivity at the final time
    with 50\% (90\%) of it being contained within 3~pc (50~pc).}
  \label{fig:alltime}
\end{figure}

Figure \ref{fig:alltime} shows the resulting \lya{} normalized
emissivities as a function of radius at several times.  We only
consider the shells within the radius range given in Table
\ref{tab:dynamic} to reduce the computation, and we have found that
the shells outside the given ranges do not contribute to the overall
emissivity.  We first start by calculating the total emissivity from
the first time interval ($t_0 \rightarrow t_1$), shown as the blue
line in the Figure.  Then at the next output time $t_2$, we calculate
the total emissivity in the next interval ($t_1 \rightarrow t_2$) and
add it to the previous profile.  This process is repeated until we
reach the final output simulation time.  In other words, at some time
$t_n$, we construct the time-integrated emissivity profile by
discretely adding each time interval:
\begin{equation}
  \label{eqn:total_eta}
  \eta^{\rm total}(t_{\rm n}, r) = \sum_i^n \eta(t_{\rm i} \rightarrow
  t_{\rm n}, r) = \sum_i^n \sum_{\rm shells} \eta^s(t_{\rm i}
  \rightarrow t_{\rm n}, r)
\end{equation}
where $n$ is an integer in the interval $[0,5]$, and its maximum
corresponds to the number of simulation outputs considered.  During
the summation, we smooth the Monte Carlo results with the kernel
density estimation (KDE) method in \texttt{scipy} with the default
parameters \citep{scipy}.  However because $n$ is small, we have
numerical artifacts at large radii from the addition of the local
maxima from previous times (i.e. $i < n$), but they do not affect the
accuracy of the final emissivity profile.

As the halo collapses, the \lya{} emissivity progressively becomes
more centrally concentrated because of the increased photon generation
rate from the higher densities.  At the final time (yellow line in
Figure \ref{fig:alltime}) when the maximum density $n_{\rm H} = 3
\times 10^{11}~\cubecm$, we see that the bulk of the \lya{} is
contained within $r < 10^{19} \unit{cm} \simeq 3 \unit{pc}$ that is
approximately the Jeans length of the central object.  At this radius,
the number density $n_{\rm H} \simeq 10^4~\cubecm$, above which the
medium becomes prone to \lya{} scattering and a reduction in radiative
cooling.  Also we show cumulative emissivity profile within a radius
$R$
\begin{equation}
  \eta(<R, t_{\rm final}) = \int_0^R \eta^{\rm total}(t_{\rm final})
  \, dV
\end{equation}
in Figure \ref{fig:alltime} as the dashed cyan line, illustrating that
50\% (90\%) of the \lya{} radiation is contained within $\sim$3~pc
(50~pc).

\begin{figure}
  \includegraphics[width=\columnwidth]{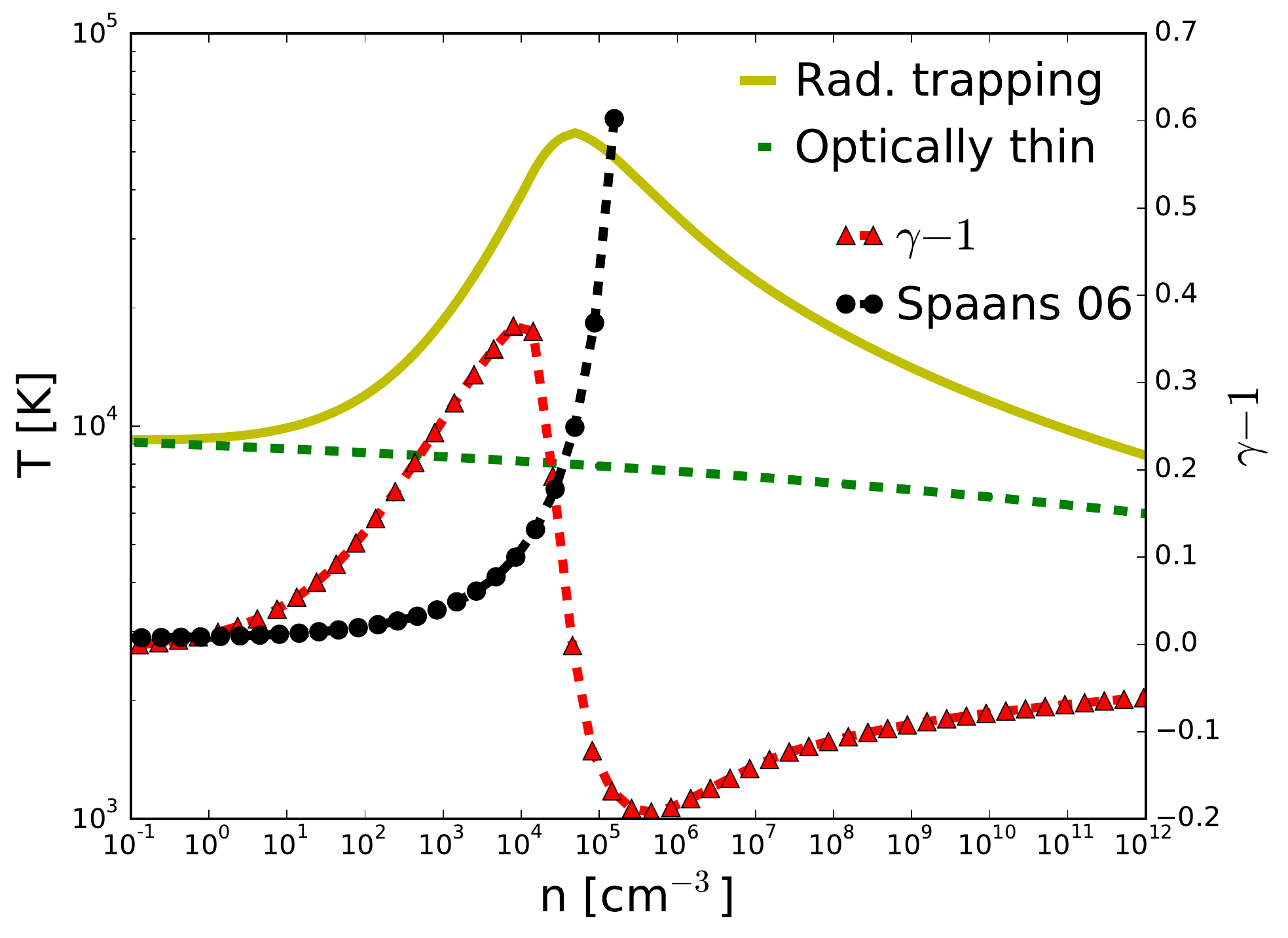}
  \caption{{\it Left axis:} Effective equation of state derived from
    the \lya{} radiation transfer calculation (solid yellow) and
    optically-thin cooling rates (green dashed).  The gas heats from
    the initial temperature of 9,000~K to 50,000~K by $n_{\rm H} = 3
    \times 10^4~\cubecm$ from \lya{} trapping and then cools to
    $10^4\unit{K}$ at higher densities.  {\it Right axis:} Adiabatic
    index $\gamma - 1$ of the effective equation of state from this
    work (red dashed) that well describes the suppressed \lya{}
    cooling from radiation trapping and from the analytical work of
    \citet[black dashed]{Spaans} that diverges above $10^5~\cubecm$
    where it should be limited to 5/3.}
  \label{fig:eos}
\end{figure}

Finally with the \lya{} emissivity profile $\eta^{\rm total}$ at the
final time $t_{\rm final}$, we can determine how much the radiative
cooling is reduced.  We first convert this profile into a function of
density by using the halo radial density profile (Figure
\ref{fig:profile}).  Then we take the difference between our Monte
Carlo radiation transport result and the optically-thin emissivity
$n^{\rm src}$ (see Equations \ref{eqn:rec} and \ref{eqn:dcol}) and
convert that into an effective heating rate as a function of density.
Figure \ref{fig:eos} depicts the resulting effective equation of state
up to a number density $n_{\rm H} = 10^{12}~\cubecm$ that is
approximately the maximum density in the simulation.  Here we take the
initial temperature at $n = 0.1~\cubecm$ to be $T = 9000 \unit{K}$.
As the gas condenses, \lya{} radiation becomes more coupled to the
neutral gas, reducing its cooling rate with respect to the
optically-thin rate, resulting in the gas gradually heating to $5
\times 10^4 \unit{K}$ at $n_{\rm H} = 3 \times 10^4~\cubecm$.  However
at higher densities (smaller radii), the gas cools back to $10^4
\unit{K}$ because the \lya{} emissivity plateaus within 3~pc,
corresponding to $n_{\rm H} = 10^4~\cubecm$ in the density profile.
This heating from partially \lya{} trapping at moderate densities is
in stark contrast to the optically-thin case, where the gas slowly
cools from 9000~K to 7000~K (green dashed line).  The gas starts to
cool because the optically thin cooling rate increases as $n_{\rm e}
n_{\rm HI}$ or $n_{\rm e} n_{\rm HII}$ for collisional excitation and
recombination, respectively, while the \lya{} emissivity from the
Monte Carlo calculation has plateaued.  The combination of this
saturation and increasing optically-thin cooling rate ultimately
results in the dense gas cooling back below $10^4\unit{K}$.  We then
differentiate this effective equation of state to obtain the adiabatic
index $\gamma = 1 + d\ln T/d\ln\rho$ for an ideal gas (red dashed line
in Figure \ref{fig:eos}).  As the gas heats, $\gamma$ increases from
unity to $\sim$4/3 at $n_{\rm H} = 10^4~\cubecm$, suddenly decreases
to $\sim$4/5 at $n_{\rm H} = 10^6~\cubecm$, and then recovers back to
unity with increasing density.

In Figure \ref{fig:eos}, we compare our equation of state to the one
analytically derived from spherical symmetry in \citet{Spaans}, shown
as a dashed black line, that has the form
\begin{equation}
  \label{eqn:spaans}
  \gamma - 1 \approx - \frac{ \tfrac{1}{2} + \tfrac{7}{18}
    Bn_1^{7/18}} {\log(Cn_1^{0.5}) + Bn_1^{7/18}},
\end{equation}
where $B \approx 0.47 \unit{cm}^{7/6}$, $C \approx 10^{-34}
\unit{cm}^{3/2}$, and $n_1$ is 100 times the number density in units
of cm$^{-3}$ \citep[see][for the motivation to boost $n_1$]{Latif}.
This result describes a smooth but quick transition from isothermal to
adiabatic ($\gamma = 5/3$), but it diverges for high number densities
and must be limited to 5/3 at high densities.  The bulk of the
increase in $\gamma$ comes between $\log(n_{\rm H}/\textrm{cm}^{-3}) =
4-5$, whereas our results starts to increase from unity around
$100~\cubecm$, only reaching $\sim$4/3 at $10^4~\cubecm$.  Our
effective equation of state is still valid for high number densities
when the cloud is optically thick, whereas the \citeauthor{Spaans}
result breaks down and must be approximated with an adiabatic equation
of state.



\section{Conclusions and Discussion}
\label{sec:4} 
We have utilized a suite of Monte Carlo \lya{} radiative transfer
calculations to study the effects of \lya{} radiation trapping in a
metal-free pre-galactic halo, which we have extracted from an AMR
cosmological simulation, using {\sc Enzo}.  In this paper, we have
quantified the delayed radiation propagation and associated reduced
radiative cooling within these objects that could be precursors of
direct collapse black holes or dense stellar clusters.  From these
calculations, we have estimated an effective equation of state for
this collapsing primordial gas.  The key results of this paper are
summarized below.
\begin{enumerate}
\item By introducing a \lya{} model, we found that the primordial
  cooling rates are reduced below 20,000~K at densities above
  100~\cubecm.  Above this temperature, cooling from spontaneous
  emission in hydrogen dominates, and below this density, the gas is
  effectively optically thin to \lya{} radiation.
\item The majority of the \lya{} photons are generated within a radius
  of $\sim$1~pc and $\sim$1~kyr before the collapse of the central
  primordial gas cloud inside of a pre-galactic atomic cooling halo.
  This gas is optically thick to \lya{} radiation, which is trapped
  within the cloud, but it eventually escapes from the cloud.  Thus,
  the optically-thin cooling rates overestimate the actual cooling
  behavior of this collapsing gaseous object.
\item When we consider a static density field, whether it be uniform
  or an isothermal profile, the \lya{} radiation outward propagation
  is delayed by resonance scattering, resulting in a emissivity radial
  profile that is well described by a Gamma distribution.  Subsonic
  inward or rotational bulk velocities allow the \lya{} photons to
  shift into the wings of the line profile but have little effect on
  reducing the amount of trapping.
\item We apply these results to a dynamically collapsing halo in a
  discrete manner, where we find that the \lya{} radiation is still
  trapped.  However, the radiative cooling rates are not fully
  suppressed with the adiabatic index rising from unity to $\sim$4/3
  at $n_{\rm H} = 10^4~\cubecm$ with the temperature increasing to
  50,000~K at the same number density.  At higher densities, the
  \lya{} emissivity saturates while cooling rates from collisional
  excitation and recombination increase as $n^2$, allowing the gas to
  cool back to 10,000~K.  This thermodynamic track results in a heated
  envelope with a cooled core that will form either a dense stellar
  cluster or a supermassive star, eventually forming a massive black
  hole seed.
\end{enumerate}

We have seen that \lya{} radiation trapping alters the thermal
properties of the collapsing system, which will change its Jeans mass,
which ultimately controls the fragmentation mass scales and resulting
collapsed object.  The Bonnor-Ebert mass \citep{Bonnor, Ebert}
considers an external pressure $P_{\rm ext}$ around an isothermal
gaseous cloud, which is given by
\begin{align}
  M_{\rm BE} &= 1.18 \frac{c_{\rm s}^4}{G^{3/2}} P_{\rm ext}^{-1/2} \;
  \Ms\\
  &\simeq 20 T^{3/2} n^{-1/2} \mu^{-2} \gamma^2 \; \Ms
\end{align}
where the second expression is calculated by setting the external
pressure to the local pressure.  Previous studies of the direct
collapse black hole pre-cursors become Jeans unstable at a
Bonnor-Ebert mass around $10^5~\Ms$ at a radius of $\sim$1 pc
\citep[e.g.][]{Bromm03, Wise, Regan}, which is approximately where we
find the primordial gas to be prone to \lya{} trapping.  We find that
the gas heats to 50,000~K at this scale that is 5--6 times larger than
the typical 8000~K temperature found in studies using optically-thin
cooling rates.  This heating increases the Bonner-Ebert mass at this
scale by an order of 10--15, which will hinder the initial collapse
until the central object can accumulate additional gas.  However after
the cloud becomes gravitationally unstable, it will cool back down to
8,000--10,000~K, resulting in a cool dense core surrounded by an
envelope that is 5 times hotter.  This additional external pressure
may drive a decrease in the Bonnor-Ebert mass at higher densities.

One shortcoming of our work is the post-processing treatment of the
\lya{} radiation transport, where the additional heating does not
affect the collapse.  In the time-dependent case, we utilized the
temperature profile from the cosmological simulation that was
calculated with the optically-thin cooling rates.  But any \lya{}
feedback will change the gas temperature and thus neutral fraction
that will ultimately alter the \lya{} radiation field.  When the gas
is heated above the optically-thin solution, the \lya{} photons will
scatter to the wings of the line profile faster and overall will have
longer mean free path.  Additionally, we have assumed spherical
symmetry, whereas in a full three-dimensional setup with coupled
\lya{} transfer anisotropic structures, such as bubbles or channels,
can form during the collapse \citep[e.g.][]{Smith15_Lya}, which could
have similar anisotropic behavior as ionizing radiation transport in
massive star formation \citep[e.g.][]{Krumholz09_Sci, Rosen}.  This
anisotropy may alter the accretion flows onto and through the
collapsing gas cloud.  For instance, \lya{} trapping may favor some
directions than others, creating warmer channels and inhibiting any
accretion through those solid angles.

Such feedback loops would create a complex interplay between accretion
flows, shocking onto the Jeans unstable gas cloud, \lya{} radiation
trapping, and the resulting thermal and hydrodynamic response.  This
will likely alter the angular momentum and entropy of the infalling
gas and could have an effect on the outcome of the collapsing object
-- the spin and mass of a direct collapse black hole, or the star
formation efficiency and size of a dense stellar cluster.  As
computational methods and hardware improve, it is becoming feasible to
perform \lya{} radiation transport coupled with the hydrodynamics to
resolve these complexities arising from the aforementioned feedback
processes \citep[e.g. see a discussion in][]{Smith17_DCBH} that will
bring us closer to resolving the nature of the initial central object
of these highly irradiated, metal-free, pre-galactic halos.


\section*{Acknowledgments}

This work was supported by National Science Foundation grants
AST-1333360 and AST-1614333, NASA grant NNX17AG23G, and Hubble theory
grants HST-AR-13895 and HST-AR-14326.  This research has made use of
NASA's Astrophysics Data System Bibliographic Services.  The
calculations were performed on XSEDE's Stampede resource with XSEDE
allocation AST-120046.  The majority of the analysis and plots were
done with \textsc{yt} \citep{yt_full_paper} and \textsc{matplotlib}
\citep{matplotlib}.  \textsc{Enzo} and \textsc{yt} are developed by a
large number of independent researchers from numerous institutions
around the world. Their commitment to open science has helped make
this work possible.




\newpage
\bibliographystyle{mnras}
\bibliography{lya}



\appendix
\section{Cooling with approximate \lya{} radiative transfer}
\label{appendix}

\subsection{Average number of scattering events}

\begin{table*}
  \centering
  \caption{Coefficients $A_{ij}$ for the exponential fit to the number
    of scattering events in Equation (\ref{eqn:Aij})}
  \label{tab:Aijtable}
  \begin{tabular*}{0.65\textwidth}{@{\extracolsep{\fill}} c | rrrrrr}
    \hline
    & $i=0$ & $i=1$ & $i=2$ & $i=3$ & $i=4$ & $i=5$\\
    \hline
    $j=0$ & 2.597(0) & 1.193(0) & 4.021(--2) & 2.375(--3) & 8.825(--5) & --6.167(--6)\\
    $j=1$ & --2.000(0) & --3.342(--1) & --9.067(--3) & --3.954(--4) & 1.008(--5) & \\
    $j=2$ & 3.400(--1) & 3.150(--2) & 7.111(--4) & 2.748(--7) & & \\
    $j=3$ & --2.422(--2) & --1.351(--3) & --1.157(--5) & & & \\
    $j=4$ & 8.436(--4) & 2.059(--5) & & & & \\
    $j=5$ & --1.139(--5) & & & & & \\
    \hline
  \end{tabular*}
  \vspace{0.5em}
  \parbox[t]{0.65\textwidth}{Note: The values are in scientific notation
    with the exponent in parentheses.}
\end{table*} 



We base our treatment of \lya{} radiation trapping in the radiative
cooling rate calculation described and presented in Sections
\ref{sec:cooling} and \ref{sec:cooling-results}, respectively, on the
\citet{Omukai} model.  In order to calculate the average number of
scattering events before escaping the system, we solve the radiative
transfer equation for \lya{} photons with the Eddington approximation
in the isotropic limit \citep{Adams2} that describes the evolution of
intensity $J$ as a function of optical depth $\tau$ and frequency
shift $x \equiv (\nu - \nu_0)/\Delta \nu_{\rm D}$ as
\begin{equation}
  \label{eq:rte}
  \frac{\nabla^2J(\tau,x)}{3H^2(a,x)} = J(\tau, x) -
  \int_{-\infty}^{\infty}J(\tau,x^{\prime}) \, q(x,x^{\prime}) \, dx^{\prime}
  - S(\tau,x),
\end{equation}
where $H$ is the normalized Voigt profile (Equation \ref{eqn:voigt}).
Recall that $a \equiv A_{12}/4\pi\Delta\nu_{\rm D}$ (Section
\ref{sec:cooling}).  Here $q(x,x^{\prime})$ is the normalized
redistribution function that describes the frequency shifts during
scattering events in the atom's rest frame \citep{Hummer}.  The source
function $S(\tau,x)$ describes the generation of \lya{} radiation at
some optical depth and frequency.  Using a Taylor expansion of the
redistribution functions \citep[e.g.][]{Adams2, Harrington, Rees}, the
radiation transfer equation can be formulated as a Poisson equation,
\begin{equation}
  \label{eqn:poisson}
  \frac{\partial^2J}{\partial\tau^2} + \nabla^2 J = -3
  \frac{S(\tau,x)}{4\pi}.
\end{equation}
The equation is solved in spherical symmetry with the boundary
condition
\begin{equation}
  \frac{\partial J(x,\tau)}{\partial \tau} = -\frac{3}{2} H(a,x)
  J(\tau,x),
\end{equation}
describing a system with $\tau = 0$ at the center and $\tau = \tau_0$
at the outer boundary $r = R$.  Considering an isotropic source at
some optical depth $\tau_{\rm s}$, the analytical solution \citep[for
the complete derivation, see Appendix C in][]{Dijkstra} to Equation
(\ref{eqn:poisson}) is,
\begin{equation}
  \label{eq:soln}
  J(\tau,\sigma) = \frac{\sqrt{6}}{16\pi^2 R} \frac{1}{\tau_0 \tau
    \tau_{\rm s}}
  \sum_{n=1}^{\infty} \sin(\lambda_{\rm n} \tau) \sin(\lambda_{\rm n}
  \tau_{\rm s}) \frac{\exp(-\lambda_{\rm n}|\sigma|)}{\lambda_{\rm n}},
\end{equation}
where $\sigma \equiv (2\pi/27)^{1/2} x^3/a$.  The values $\lambda_{\rm
  n}$ are the coefficients in the solution to the equation,
\begin{equation}
  \frac{d^2J}{d\tau^2} + \lambda^2 J = 0
\end{equation}
that has solutions in the form $J_n = A\cos(\lambda_{\rm n}\tau)$
\citep{Unno, Harrington}.  After determining the values of $\lambda$
and thus the solution to $J$, it can be integrated from the center to
the optical depth $\tau_s$ and compared to the intensity $J$ at the
outer boundary.  The respective ratio of these two quantities relates
the number $N_{\rm sc}$ of scatterings inside a sphere with optical
depth $\tau_s$ to the total number $N_{\rm bd}$ of photons emitted
from boundary $\tau_0$
\begin{align}
  \frac{N_{\rm sc}(x,\tau_s)}{N_{\rm bd}(x,\tau_s)} 
  &= \frac{\int_0^{\tau_0} 4\pi\tau^2 J(\tau,\sigma) \, d\tau}
    {J(\tau_0,\sigma)}\\
  &= \frac{\mathcal{L}(x,\tau_0,\tau_{\rm s}) / [ 2i \tau_0
    A^2(x,\tau_0) ] } 
    {  \sin(\pi\tau_{\rm s}/\tau_0) / \{[3H(a,x)] [\cos(\pi\tau_{\rm s}/\tau_0) + \cosh(\sigma/\tau_0)]\} },\nonumber
\end{align}
where 
\begin{equation}
  \mathcal{L}(x, \tau_0, \tau_{\rm s}) 
  \equiv \textrm{Li}_2[A(x,\tau_0) \, (|\sigma| - i\tau_{\rm s})] -
  \textrm{Li}_2[-A(x,\tau_0) \, (|\sigma| + i\tau_{\rm s})]
\end{equation}
and
\begin{equation}
  A(x,\tau_0) = \frac{\pi}{\tau_0} \left(1-\frac{2}{3H(a,x)\tau_0+2} \right).
\end{equation}
The function $\textrm{Li}_{n}(z) \equiv \sum_{k=1}^\infty (z^k / k^n)$
is an $n=2$ polylogarithmic function defined in the complex plane.
From this solution, we can integrate over the frequency $x$ (or its
equivalent $\sigma$) and optical depth $\tau_s$ from the center to the
boundary, determining the average number of scatterings for a photon
escaping from the system to be
\begin{equation}
  N_{\rm esc} = \frac{3}{\tau_0^3} \int_0^{\tau_0}
  \frac{\int_{-\infty}^{\infty} N_{\rm sc}(x,\tau_{\rm s}) \, dx} 
  {\int_{-\infty}^{\infty} N_{\rm bd}(x,\tau_{\rm s}) \, dx} \tau_{\rm s}^2 \,
  d\tau_{\rm s}.
\end{equation}  
This expression for the average number $N_{\rm esc}$ of scatterings
cannot be solved analytically, so we numerically integrate it for 1600
equally log-spaced pairs of temperature $T$ in the range of $10^3 -
10^9 \unit{K}$ (corresponding to some $\Delta\nu_{\rm D}$) and optical
depth $\tau_0$ in the range $10^3 - 10^6$.  We fit these numerical
results to the two-variable exponential polynomial function
\begin{equation}
  \label{eqn:Aij}
  \frac{N_{esc}(T,\tau_0)}{\tau_0} = \sum_{i,j}^{i+j \le 5} \exp
  \left[ A_{\rm ij}  \left(\ln T\right)^i  \left(\ln \tau_0\right)^j \right]
\end{equation}
%
with the coefficients $A_{\rm ij}$ given in Table \ref{tab:Aijtable}.

\subsection{Radiation emission in a two-level system}

We simplify the \lya{} emission process by considering a two-level
system because the spontaneous transitions from more excited states
reside in the optically thin regime \citep{Shang}.  The number density
of the first excited state is related to the ground state by
\begin{equation}
  \label{eqn:n2n1}
  \frac{n_2}{n_1} = \frac{C_{12} + (g_2/g_1) u_{\nu} A_{21}} 
  {C_{21} + (1+u_{\nu}) A_{21}},
\end{equation}
where $C_{\rm ij} = n_{\rm e} k_{\rm ij}(\textrm{e}) + n_{\rm H}
k_{\rm ij}(\textrm{H})$ is the collisional de-excitation rate by free
electrons and hydrogen atoms, and $g_n = 2n^2$ is the statistical
weight \citep[e.g.][]{Tielens85, Omukai}.  The quantity
\begin{equation}
  u_\nu = \frac{c^2}{2hv_{21}^3} J_{\rm cont}(\nu_{21}) \gg 1
\end{equation}
is related to the incoming photon flux at the energy difference
$E_{21} = h\nu_{21} = 10.2 \unit{eV}$ between the states.  The
population density of the excited state (Equation \ref{eqn:n2n1}) will
change through collisional processes and spontaneous emission
\begin{equation}
  \frac{dn_{2 \rightarrow 1}}{dt} = k_{21}(\textrm{e}) n_2 n_{\rm e} +
  k_{21}(\textrm{H}) n_2 n_1 + A_{21} n_2,
\end{equation}
and the associated cooling rate per unit volume is reduced by the
number of scatterings (Equation \ref{eqn:Aij}),
\begin{equation}
  \Lambda_{2 \rightarrow 1} = \frac{h\nu_{21}}{N_{\rm esc}}
  \frac{dn_{2 \rightarrow 1}}{dt}.
\end{equation}
We use the collisional coefficient rates from \citet{Omukai}:
\begin{equation}
  k_{21}(\textrm{e}) = 1.155 \times 10^{-8}
  \frac{\sqrt{\beta(\beta+1)}} {\beta+0.28} \unit{cm}^3 \unit{s}^{-1},
\end{equation}
\begin{equation}
  k_{21}(\textrm{H}) = 1.454 \times 10^{-15} \frac{T^{1/2} + 1.693
    \times 10^{-5} \, T^{3/2}}{1 + 8.46 \times 10^{-17} \, T^2}
  \unit{cm}^3 \unit{s}^{-1},
\end{equation}
where $\beta \equiv E_{21}/kT$.  In the temperature range $T = 8-10
\times 10^{3} \unit{K}$, radiation originating from the two-photon
process is in the optically thin regime, but we need to consider it in
the model to obtain accurate electron states for the first excited
state.  The ratio between the $2s$ and $2p$ states is given by
\begin{equation}
  \frac{n_{\rm 2s}}{n_{\rm 2p}} = \frac{g_{\rm 2s}}{g_{\rm 2p}}
  \frac{C_{\rm 2s2p}}{C_{\rm 2s2p} + A_{\rm 2s1s}},
\end{equation}
where $(g_{\rm 2s}, g_{\rm 2p}) = (2,6)$, and $A_{\rm 2s1s} = 8.23
\unit{s}^{-1}$.  The collision rate can be described with the fit
\citep{Omukai}
\begin{equation}
  C_{\rm 2s2p} = 6.21 \times 10^{-4} T^{-1/2} \ln(5.7T) \left[1 +
    \frac{0.78}{\ln(5.7T)} \right] n(\textrm{e}) \unit{s}^{-1}.
\end{equation}
Lastly, the Einstein A-coefficient associated with spontaneous
emission for the 2p $\rightarrow$ 1s and 2s $\rightarrow$ 1s
transitions are respectively
\begin{equation}
A_{21} = \frac{n_{\rm 2p}}{n_2}A_{\rm 2p1s}
\quad\textrm{and}\quad
A_{\rm 2ph} = \frac{n_{\rm 2s}}{n_2}A_{\rm 2s1s},
\end{equation}
resulting in the cooling rate from the two-photon process
\begin{equation}
  \Lambda_{\rm 2ph} = h\nu_{21} A_{\rm 2ph} n_{\rm 2s}
\end{equation}



\bsp 
\label{lastpage}
\end{document}